%% file: main.tex
\documentclass[conference]{IEEEtran}

\usepackage{cite}
\usepackage{float}
\usepackage{xcolor}

\makeatletter
\def\@IEEEsectpunct{\ }
\makeatother
\usepackage{microtype}
\usepackage{graphicx}
\usepackage{booktabs}
\usepackage{fancyvrb}
\usepackage{listings}
\usepackage{amsmath}
\usepackage{amssymb}
\usepackage{enumitem}
\usepackage[hyperfootnotes=false]{hyperref}
\usepackage{url}
\usepackage{tikz}
\usetikzlibrary{arrows.meta,decorations.pathmorphing}
\usepackage{pifont}
\hypersetup{hypertexnames=false}
\hypersetup{
  colorlinks=true, linkcolor=red!60!black, citecolor=blue!60!black,
  urlcolor=blue!60!black,
  pdftitle={A Fast, Programmable GPU-to-CPU Invocation Layer for GPU-Initiated Communication},
  pdfauthor={Jiakun Yan, Shiksha Patel, David Sidler, Gengyu Rao, Brandon Potter, Bradford Beckmann},
  pdfkeywords={GPU communication, device-initiated communication, reverse offload, GPU-to-CPU callbacks}}

\floatstyle{plain}
\newfloat{listing}{tbp}{lol}
\floatname{listing}{Listing}

\input{macros}

\lstdefinestyle{syscode}{
  language=C++,
  basicstyle=\ttfamily,
  columns=fullflexible,
  keepspaces=true,
  showstringspaces=false,
  aboveskip=0pt,
  belowskip=0pt,
  morekeywords={__global__,__device__,parallel_for,workgroup,wave,each,in,loop,
    u32,u64,uint8_t,uint32_t,uint64_t,uintptr_t,size_t},
  keywordstyle=\color{blue!70!black},
  commentstyle=\color{green!45!black},
}

\title{A Fast, Programmable GPU-to-CPU Invocation\\Layer for GPU-Initiated Communication}

\author{\IEEEauthorblockN{Jiakun Yan, Shiksha Patel, David Sidler, Gengyu Rao, Brandon Potter, Bradford Beckmann}
\IEEEauthorblockA{Advanced Micro Devices, Inc.\\
\{jiakun.yan, shiksha.patel, david.sidler, lorri.rao, brandon.potter, brad.beckmann\}@amd.com}}

\begin{document}
\bstctlcite{IEEEexample:BSTcontrol}

\maketitle

\begin{abstract}
\input{sections/abstract}
\end{abstract}


\input{sections/introduction}

\input{sections/related_work}
\input{sections/design}

\input{sections/queue}

\input{sections/p2p}

\input{sections/ep}
\input{sections/evaluation}
\input{sections/conclusion}

\section*{Acknowledgment}

AMD, the AMD Arrow logo, AMD Instinct, Infinity Fabric, ROCm, and combinations thereof are trademarks of Advanced Micro Devices, Inc. Other product names used in this publication are for identification purposes only and may be trademarks of their respective companies.

\bibliography{refs}
\bibliographystyle{IEEEtran}

\end{document}

%% file: macros.tex
\usepackage{xspace}

\newcommand{\sys}{KIWI\xspace}
\newcommand{\sysptp}{KIWI-P2P\xspace}
\newcommand{\sysep}{KIWI-EP\xspace}

\usepackage{xcolor}
\newif\ifshowcomments
\showcommentstrue

%% file: sections/abstract.tex
Modern GPU workloads increasingly rely on device-initiated communication to let GPU kernels communicate directly without synchronizing with the host at kernel boundaries. Existing libraries commonly favor placing transport control on GPU threads, an approach known as GPUDirect Async (GDA), and treat reverse offload to host proxy threads as a slower fallback. We argue that reverse offload can outperform GDA and should be a first-class communication path. We further propose exposing reverse offload as a programming interface that lets users define workload-specific communication operations.

We present \sys, a programmable reverse-offload framework that achieves sub-microsecond one-way device-to-host latency and hundreds of millions of operations per second. We address practical challenges of scaling reverse offload, including contention on shared queues, polling overhead, and load imbalance across proxies. \sysptp, a SHMEM-like library built on \sys, outperforms state-of-the-art GDA implementations in both latency and bandwidth. More importantly, \sysep redesigns low-latency mixture-of-experts dispatch and combine kernels around workload-defined operations, lowering expert-parallel communication latency by up to 2.3$\times$ over the best existing backend on AMD GPUs and time per output token in vLLM by up to 29\%. The source code of \sys is available at \url{https://github.com/AMDResearch/kiwi}.

%% file: sections/introduction.tex
\section{Introduction}
\label{sec:introduction}

As large language models and their infrastructure evolve rapidly, new techniques such as fused kernels, persistent kernels, and mixture-of-experts (MoE) communication kernels continue to emerge~\cite{comet2025,fusedcollectives2024,mpk2026,deepep2025}. These workloads require more dynamic and flexible communication than traditional NCCL-style, host-initiated collectives provide. Device-initiated GPU communication libraries, such as NVSHMEM and rocSHMEM, have therefore become increasingly popular~\cite{nvshmem2020,rocshmem2020}.

Early device-initiated GPU communication relied on reverse offload to host proxy threads, as represented by NVSHMEM's IBRC transport. Later systems sought to remove the host from the control path by allowing GPU threads to interact directly with the network interface card (NIC). This approach, commonly known as GPUDirect Async (GDA), promised to eliminate proxy-throughput bottlenecks and reduce control-path latency~\cite{gpudirectasync2022,deepseekv3technicalreport}. Most device-initiated libraries now provide a GDA transport. As GDA has gained adoption, reverse offload was increasingly relegated to a fallback path and commonly perceived as the lower-performance option.

However, GDA has inherent limitations. GPU cores favor bulk-parallel computation and data movement, whereas the NIC-facing control path is a sequence of dependent scalar steps of protocol selection, descriptor enqueueing, and doorbell ringing. On a GPU, these steps require branches, global-memory accesses, and system-scope fences. The weak sequential performance of a GPU thread thus limits the communication primitives and patterns that GDA can express efficiently. Moreover, a GPU furnishes far more parallel execution resources than a NIC provides independent communication endpoints. A GDA runtime therefore allocates many NIC endpoints, such as InfiniBand queue pairs, to reduce contention, yet GPU threads still outnumber them and must synchronize to share them. Both synchronization and large endpoint counts limit GDA scalability.

\input{figures/gda_vs_ro}

In this paper, we advocate revisiting reverse offload for device-initiated GPU communication. Reverse offload keeps initiation on the GPU but moves the NIC-facing control path to CPU cores (Figure~\ref{fig:gda-vs-ro}), which efficiently execute serial control flow while the GPU continues useful computation. It also decouples GPU concurrency from transport concurrency: many GPU producers submit work through lightweight queues in host memory to a smaller pool of CPU workers, each of which drives a NIC endpoint, so the number of NIC endpoints no longer has to grow with the number of GPU threads. We show that a well-optimized reverse-offload software stack can outperform state-of-the-art GDA implementations in both latency and bandwidth.

We further argue that the reverse-offload stack should be open to kernel developers. GPU kernels often need operations beyond the off-the-shelf primitives of vendor communication libraries, such as multicast or collectives with customized completion and progress semantics. Composing these operations from standard SHMEM primitives introduces redundant control-path overhead, and adding them to an existing vendor library requires modifying its internal transport layers, preserving its correctness and performance guarantees, and carrying the change across every new release. With programmable reverse offload, a new operation is instead a self-contained host callback in the application's own code, improving both flexibility and performance.

\input{figures/overview}

We present \sys (Kernel-Initiated, Workload-Defined Invocations), a programmable reverse-offload framework that reinvents the traditional proxy transport from a minimalistic perspective. \sys provides a minimal core interface that enables users to register arbitrary host functions as callbacks and invoke them asynchronously from GPU kernels (Figure~\ref{fig:sys-overview}). To make it efficient, \sys uses a new single-producer, single-consumer (SPSC) queue that publishes each request with one wave-wide\footnote{A wave (a.k.a.\ wavefront) is AMD's counterpart of an NVIDIA warp.} store over PCIe\footnote{We assume a PCIe connection between the CPU and the GPU; the design also applies to other CPU--GPU links, such as AMD's Infinity Fabric\texttrademark{}.} and no atomic operations or memory fences, achieving sub-microsecond one-way latency. On the host, polling an idle queue costs a single load that typically hits in the L1 cache, so a few proxy threads can monitor many queues without becoming the bottleneck.

We evaluate \sys in two steps. We first validate it with \sysptp, which uses \sys as a conventional reverse-offload transport: it combines \sys with the Lightweight Communication Interface (LCI)~\cite{lci2025}, a host-side HPC communication library, to provide fixed SHMEM-like operations in fewer than 1,000 lines of C++. Even used this way, it achieves lower latency and higher bandwidth than state-of-the-art GDA implementations, confirming that reverse offload can be the faster path.

Building on this result, we then show what \sys's efficient and programmable reverse-offload stack brings to a real-world workload. \sysep redesigns low-latency MoE dispatch and combine kernels for expert parallelism (EP) around three workload-defined operations: a multicast put, a count all-to-all, and a barrier, each a single invocation from the GPU. This moves the communication-protocol control-flow complexity to the host and streamlines the GPU kernels into packing, posting, and unpacking data. \sysep lowers dispatch and combine latency by up to 2.3$\times$ over the best existing backend on AMD GPUs; integrated into vLLM serving GLM-5.3, it lowers time per output token (TPOT) by up to 29\%.

This paper makes the following contributions:
\begin{itemize}[leftmargin=*,nosep]
  \item We propose reverse offload as a programming model for device-initiated communication: instead of calling a fixed set of library-defined device operations, kernel developers define their own operations as host callbacks and invoke them from GPU kernels.
  \item We realize this model in \sys, with an interface for registering arbitrary host callbacks and a queue that publishes each invocation in one wave-wide store without atomic operations (Section~\ref{sec:sys}).
  \item With \sysptp, a prototype library built on this model, we show that reverse offload outperforms state-of-the-art GDA transports in both latency and bandwidth, challenging the assumption that GDA is the faster path (Section~\ref{sec:sys-p2p}).
  \item Most importantly, with \sysep we show that programmable reverse offload pays off in production serving: EP kernels built on workload-defined operations lower latency by up to 2.3$\times$ over existing EP stacks and vLLM time per output token by up to 29\% (Section~\ref{sec:sys-ep}).
\end{itemize}

%% file: figures/gda_vs_ro.tex
\begin{figure}[t]
\centering
\begin{tikzpicture}[
    font=\footnotesize,
    every node/.style={align=center},
    dev/.style={draw, rounded corners=2pt, minimum height=5.5mm, minimum width=10mm, inner sep=1pt},
    ctrl/.style={-{Stealth[length=1.8mm]}, thick},
    dat/.style={-{Stealth[length=2.4mm, width=2.4mm]}, line width=1.4pt, dash pattern=on 3pt off 2pt, orange!80!black},
    title/.style={font=\footnotesize\bfseries},
    sub/.style={font=\scriptsize},
]
\foreach \x/\name/\ro in {0/{(a) GDA}/0, 4.25/{(b) Reverse offload}/1} {
  \begin{scope}[xshift=\x cm]
    \node[title] at (1.9, 0.15) {\name};
    \node[dev] (ga) at (1.75, -0.4) {GPU A};
    \ifnum\ro=0
      \node[dev, minimum width=9mm, draw=gray!45, text=gray!55] (ca) at (0.45, -1.2) {CPU A};
    \else
      \node[dev, minimum width=9mm] (ca) at (0.45, -1.2) {CPU A};
    \fi
    \node[dev] (na) at (1.75, -2.0) {NIC A};
    \node[dev] (gb) at (3.35, -0.4) {GPU B};
    \node[dev] (nb) at (3.35, -2.0) {NIC B};
    \draw[dat] ([xshift=2.5mm]ga.south) -- ([xshift=2.5mm]na.north);
    \draw[dat] (na) -- (nb);
    \draw[dat] (nb) -- (gb);
    \ifnum\ro=0
      \draw[ctrl] ([xshift=-2.5mm]ga.south) -- ([xshift=-2.5mm]na.north);
    \else
      \draw[ctrl] (ga.west) -| (ca.north);
      \draw[ctrl] (ca.south) |- (na.west);
    \fi
  \end{scope}
}
\draw[ctrl] (1.6, -2.75) -- (2.2, -2.75) node[right, sub, text=black] {control path};
\draw[dat]  (4.6, -2.75) -- (5.2, -2.75) node[right, sub, text=black] {data path};
\end{tikzpicture}
\caption{GDA versus reverse offload. (a) With GDA, GPU threads drive the NIC themselves. (b) With reverse offload, the GPU hands each request to a CPU proxy, which drives the NIC. In both, the NIC moves payload data directly between GPU memories.}
\label{fig:gda-vs-ro}
\end{figure}
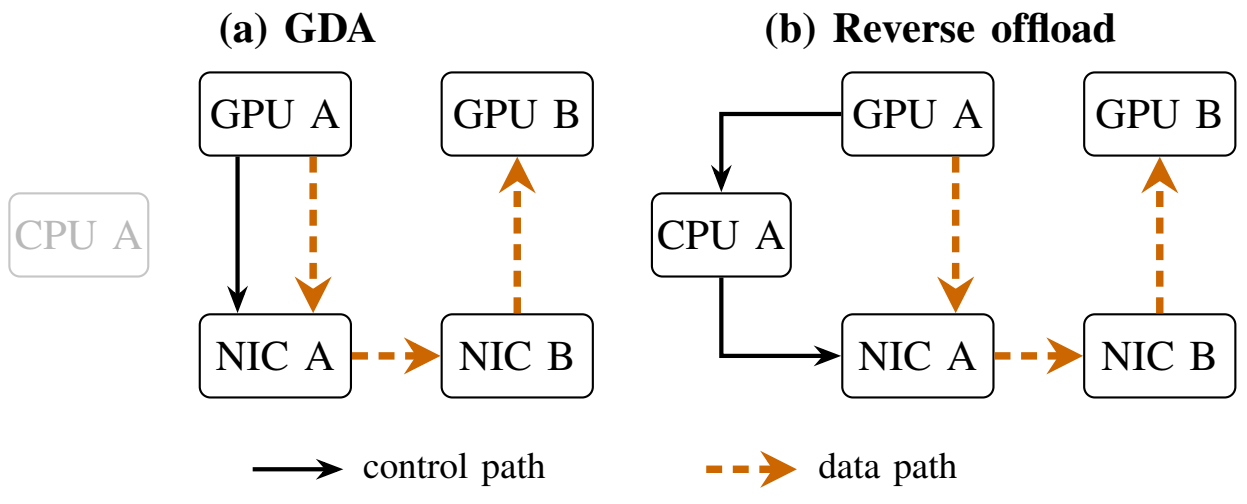

%% file: figures/overview.tex
\begin{figure}[t]
\centering
\begin{tikzpicture}[
    font=\footnotesize,
    every node/.style={align=center},
    dev/.style={draw, rounded corners=2pt, minimum height=5.5mm, inner sep=1pt},
    ctrl/.style={-{Stealth[length=1.8mm]}, thick},
    q/.style={-{Stealth[length=1.5mm]}, semithick},
    thr/.style={decorate, decoration={snake, amplitude=0.5mm, segment length=1.6mm}, semithick},
    title/.style={font=\footnotesize\bfseries},
    sub/.style={font=\scriptsize},
]
\colorlet{rt}{gray!30}      %
\definecolor{sysblue}{HTML}{0072B2}  %
\colorlet{kw}{sysblue!22}  %
\colorlet{hl}{violet!20}    %
\newcommand{\layer}[5]{\fill[#5] (#1, #3) rectangle (#2, #4);}
\begin{scope}[xshift=1.6cm]
  \node[title] at (0, 0.15) {(a) Traditional reverse offload};
  \node[sub, rotate=90, text=black!70] at (-1.78, -0.65) {GPU};
  \node[sub, rotate=90, text=black!70] at (-1.78, -2.5) {CPU};
  \layer{-1.6}{1.6}{-0.2}{-0.65}{white}
  \layer{-1.6}{1.6}{-0.65}{-1.1}{rt}
  \node[sub] at (0, -0.425) {GPU kernel};
  \node[sub] at (0, -0.875) {\texttt{put}, \texttt{get}, \texttt{quiet}, \ldots};
  \draw[rounded corners=2pt] (-1.6, -0.2) rectangle (1.6, -1.1);
  \foreach \s in {-1.3, -0.8, -0.3, 0.3, 0.8, 1.3}
    \draw[semithick] (\s, -1.1) -- (0, -1.4);
  \draw[-{Triangle[length=1.8mm, width=4.2mm]}, line width=2.4mm, gray!55] (0, -1.36) -- (0, -1.8);
  \node[sub, right] at (0.25, -1.58) {shared queue};
  \layer{-1.6}{1.6}{-1.8}{-3.2}{rt}
  \draw[thr] (0.75, -1.85) -- (0.75, -3.15);
  \node[sub, anchor=west] at (-1.5, -2.5) {built-in proxy};
  \draw[rounded corners=2pt] (-1.6, -1.8) rectangle (1.6, -3.2);
  \draw[ctrl] (0, -3.2) -- (0, -3.5);
  \node[dev, fill=rt, minimum width=32mm] at (0, -3.77) {NIC};
\end{scope}
\begin{scope}[xshift=5.5cm]
  \node[title] at (0, 0.15) {(b) \sys};
  \node[sub, rotate=90, text=black!70] at (-1.78, -0.65) {GPU};
  \node[sub, rotate=90, text=black!70] at (-1.78, -2.5) {CPU};
  \layer{-1.6}{1.6}{-0.2}{-0.65}{white}
  \layer{-1.6}{1.6}{-0.65}{-1.1}{kw}
  \node[sub] at (0, -0.425) {GPU kernel};
  \node[sub] at (0, -0.875) {\texttt{invoke(cb, args)}};
  \draw[rounded corners=2pt] (-1.6, -0.2) rectangle (1.6, -1.1);
  \foreach \s in {-1.4, -1.15, -0.9, 0.9, 1.15, 1.4}
    \draw[q] (\s, -1.1) -- (\s, -1.8);
  \node[sub] at (0, -1.45) {per-wave queues};
  \layer{-1.6}{1.6}{-1.8}{-2.27}{kw}
  \layer{-1.6}{1.6}{-2.27}{-2.73}{white}
  \layer{-1.6}{1.6}{-2.73}{-3.2}{hl}
  \node[sub, anchor=west] at (-1.5, -2.035) {progress};
  \node[sub, anchor=west] at (-1.5, -2.5) {callbacks};
  \node[sub, anchor=west] at (-1.5, -2.965) {host library};
  \foreach \s in {0.2, 0.55, 0.9, 1.25}
    \draw[thr] (\s, -1.85) -- (\s, -3.15);
  \draw[rounded corners=2pt] (-1.6, -1.8) rectangle (1.6, -3.2);
  \draw[ctrl] (0, -3.2) -- (0, -3.5);
  \node[dev, fill=hl, minimum width=32mm] at (0, -3.77) {NIC};
\end{scope}
\foreach \c/\t/\lx in {white/user code/0.0, rt/library runtime/1.75, kw/\sys/4.05, hl/host library/5.15} {
  \draw[fill=\c] (\lx, -4.4) rectangle ++(0.3, 0.2);
  \node[sub, right] at (\lx+0.3, -4.3) {\t};
}
\end{tikzpicture}
\caption{Traditional reverse offload versus \sys. (a) Traditional reverse offload hides the proxy and the NIC behind a fixed device API; all GPU threads share one queue, typically served by a single proxy thread. (b) \sys exposes a single call, \texttt{invoke}, that runs user-defined callbacks on the CPU. Each wave owns its queue, several proxy threads poll the queues, and callbacks drive the NIC through any host library, such as LCI.}
\label{fig:sys-overview}
\end{figure}
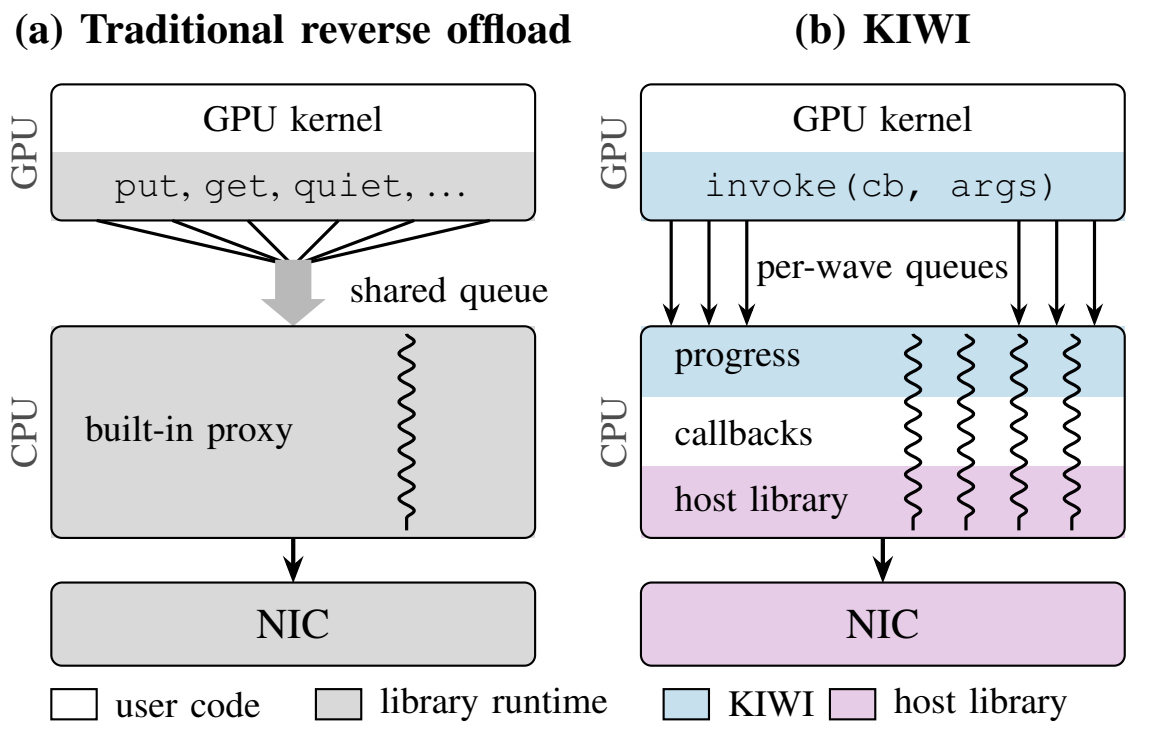

%% file: sections/related_work.tex
\section{Related Work}
\label{sec:related}

Host communication libraries such as MPI and OpenSHMEM have long served HPC applications~\cite{mpi50,openshmem16}; \sysptp and \sysep use one such library, LCI~\cite{lci2025}, as their host transport. GPU collective libraries such as NCCL, RCCL, oneCCL, and NCCLX bring a similar host-initiated interface to GPU buffers~\cite{nccl2026,demystifyingnccl2025,rccl2026,oneccl2026,ncclx2025}. More recently, device-initiated communication libraries let GPU kernels directly issue communication operations. NVSHMEM introduced OpenSHMEM-style operations callable from GPU kernels~\cite{nvshmem2020}, followed by rocSHMEM and Intel SHMEM~\cite{rocshmem2020,intelshmem2024}; MSCCL++ exposes device-side channel primitives~\cite{mscclpp2026}, and NCCL and RCCL now provide device APIs through GPU-Initiated Networking (GIN)~\cite{ncclgin2025,rcclgin2026}.

Expert-parallel (EP) dispatch and combine are among the most demanding users of device-initiated libraries. Earlier MoE systems implement them with host-initiated all-to-all collectives~\cite{deepspeedmoe2022,tutel2023}. DeepEP introduced low-latency device-initiated EP kernels on NVSHMEM's GDA transport~\cite{deepep2025}, and many EP libraries have since followed~\cite{hybridep2026,mooncakeep2025,nixl2025,pplxkernels2025,swiftep2026,ncclep2026}, including ROCm\texttrademark{} DeepEP and MORI-EP on AMD GPUs~\cite{rocmdeepep2025,mori2025,moriep2025}.

Device-initiated libraries implement the NIC-facing control path in one of two ways. Reverse offload uses CPU proxy threads to translate descriptors generated on the GPU into network operations; most of the libraries above provide such a path. GDA instead lets GPU threads build work requests and ring the NIC doorbell themselves. Early work found that GPU threads post work requests far more slowly than CPU threads~\cite{oden2014ibverbs,gpurdma2016} and worked around this with CPU-prepared operations or batching across warps~\cite{gpudirectasync2018,gputn2017,gpucentric2017}. NVSHMEM's IBGDA transport later brought GPU-built work requests into production~\cite{gpudirectasync2022}, and NVSHMEM, rocSHMEM, NCCL GIN, and MORI now all provide GDA transports~\cite{rocshmem2026,ncclgin2025,mori2025}. Recent work, including UCCL-EP and fabric-lib, shows that a well-tuned CPU proxy can match or even outperform GDA in EP kernels~\cite{uep2026,fabriclib2026}, and \sysptp shows that this also holds for general point-to-point primitives on AMD GPUs.

Existing reverse-offload paths, however, are tightly coupled to their libraries and expose a fixed set of operations. Most of their GPU-to-CPU channels are also shared by many GPU producers, which claim each slot with an atomic operation or a lock; fabric-lib avoids a queue only by restricting the GPU to fixed per-phase completion flags~\cite{fabriclib2026}. Outside communication, prior work has used GPU-to-CPU callbacks and RPC for infrequent services such as system calls, file I/O, and printing~\cite{gpu_callbacks2010,gpufs2013,genesys2018,openmpreverseoffload2023,llvmlibcgpu2023}. \sys makes programmable reverse offload practical on the performance-critical communication path by making each GPU-to-CPU handoff cheap and scalable to many GPU producers and proxy threads.

%% file: sections/design.tex
\section{KIWI Design}
\label{sec:sys}

This section presents \sys (Figure~\ref{fig:sys-overview}), a programmable reverse-offload layer that lets applications register arbitrary host functions as callbacks and invoke them efficiently from GPU kernels. \sys has two design goals, flexibility and efficiency. For flexibility, its callback-based interface lets each application define its own device-initiated operations; for efficiency, a device-to-host queue minimizes the GPU-to-CPU handoff cost with a single wave-wide store over PCIe and no locks or atomic operations on the critical path. \sys itself is independent of any communication library. Section~\ref{sec:sys-p2p} shows how to compose it with an existing host communication library to provide device-initiated communication.

\subsection{A Programmable Reverse-Offload Interface}

\sys exposes reverse offload through three steps: the host registers callbacks, GPU waves invoke them, and host threads drive their execution. Listing~\ref{lst:sys-invoke} sketches this programming model.

\begin{listing}[t]
\centering
\begin{minipage}{0.98\columnwidth}
\scriptsize
\begin{lstlisting}[style=syscode]
// Host: register a workload-defined operation.
Context ctx(num_endpoints);
cb = ctx.register_callback(
       [](Args a) { /* any host code */ });

// Host: each proxy thread runs invoked callbacks.
for each proxy thread t:
  loop: for ep in endpoints_owned_by(t):
          ctx.progress(ep);

// GPU: each wave invokes through its own endpoint.
__global__ void kernel(DeviceContext* dctx, ...) {
  h = dctx->get_device_handle(my_wave_id());
  ...
  while (!h.invoke(cb, args)) {}  // retry if full
}
\end{lstlisting}
\end{minipage}
\caption{The \sys programming model.}
\label{lst:sys-invoke}
\end{listing}

A \sys context owns a callback registry and a set of \emph{endpoints}. An endpoint is a one-way GPU-to-CPU channel backed by an SPSC queue (Section~\ref{sec:sys-queue}). Registering a callback assigns it a small integer identifier. Inside a kernel, a wave obtains a handle to an endpoint and calls \texttt{invoke} with a callback identifier and the callback's arguments; \sys serializes the arguments into a descriptor and publishes it to the endpoint. The \texttt{invoke} method is asynchronous: it returns once the descriptor is published, without waiting for the callback to run. It is also nonblocking: if the endpoint is full, it returns false immediately, and the kernel can retry, as the \texttt{while} loop in Listing~\ref{lst:sys-invoke} does, or do other useful work before trying again. On the host, \sys creates no hidden threads. The application calls \texttt{progress} on an endpoint from one of its own threads; each call pops the published descriptors and runs the corresponding callbacks.

Real kernels invoke callbacks from many waves, so the application must decide which wave pushes to each endpoint and which host proxy thread polls it. Because we use the SPSC queue for each endpoint, \sys requires that each endpoint have only one producer wave and be polled by one proxy thread at a time. In the typical mapping (Listing~\ref{lst:sys-invoke}), each wave owns one endpoint, and each proxy thread owns a disjoint set of endpoints, so neither side needs locks or atomic operations. GPU waves far outnumber host threads, so each proxy polls many endpoints. We expect this to be cheap because polling an idle endpoint costs a single load that typically hits in the proxy's L1 cache (Section~\ref{sec:sys-queue}). As a result, a few host threads can efficiently serve hundreds of waves. The application controls this mapping; \sysep, for example, adjusts the range of endpoints its proxies poll at run time (Section~\ref{sec:sys-ep}).

\paragraph{Kernels in other GPU languages.}\label{sec:sys-frontends}
The device side of \sys consists only of loads and stores that follow its queue layout, so it is not tied to HIP C++. We provide device-side libraries for FlyDSL~\cite{flydsl2026} and Triton~\cite{triton2019} that publish the same descriptors as the C++ library. Kernels written in these languages can therefore invoke the same C++ host callbacks, with no changes to the host code.

%% file: sections/queue.tex
\subsection{An Efficient Device-to-Host Queue}
\label{sec:sys-queue}

Programmable reverse offload is practical only if the GPU-to-CPU handoff is both cheap and flexible. Existing queues claim slots with atomic operations (e.g., in NVSHMEM and MSCCL++~\cite{nvshmem2020,mscclpp2026}), which are expensive. Flag-based handoffs (e.g., in fabric-lib~\cite{fabriclib2026}) are cheap, but a flag carries no description of the work, so developers must hard-code the host action for each flag. \sys's handoff combines the cost of a flag with the flexibility of a queue.

\sys provides several GPU--CPU queue types, including a multi-producer, single-consumer (MPSC) device-to-host queue; its endpoints use the SPSC device-to-host queue by default, which we describe here.

\input{figures/queue_ring}

Figure~\ref{fig:sys-queue-ring} shows the queue: a ring of 8-byte slots in GPU-accessible host memory, each pairing 4 bytes of a descriptor with a 4-byte signal. The handoff follows the low-latency (LL) protocol of GPU communication libraries~\cite{demystifyingnccl2025}: each lane (GPU thread) writes two adjacent slots with a single 16-byte store over PCIe. Because each slot reaches host memory atomically, a valid signal guarantees that the descriptor bytes beside it have arrived, so the producer needs no separate flag write or fence. With 64 lanes, one wave-wide store thus publishes a descriptor of up to 512 bytes. The producer index, a cached copy of the consumer index, and the ring metadata stay in the wave's registers, so the fullness check and producer index update are register arithmetic; the producer reads the consumer index over PCIe only when its cached copy indicates that the ring may be full. On the common path, therefore, the only significant cost of a push is the wave-wide store over PCIe.

The pop path mirrors the push path on the host. The consumer thread polls the signal of the first slot at its consumer index; once the signals of all the descriptor's slots are valid, it copies the descriptor out and advances the consumer index. Because an idle consumer always polls the same address, idle polls typically hit in the CPU's L1 cache.

Invoking a callback reuses this path and adds only argument (de)serialization and callback lookup. Section~\ref{sec:evaluation} confirms that the wave-wide store over PCIe dominates the handoff cost: the queue's round trip is within 0.1~\(\mu\mathrm{s}\) of a raw wave-wide store (Table~\ref{tab:queue-baselines}), and the programmable interface adds negligible overhead.

%% file: figures/queue_ring.tex
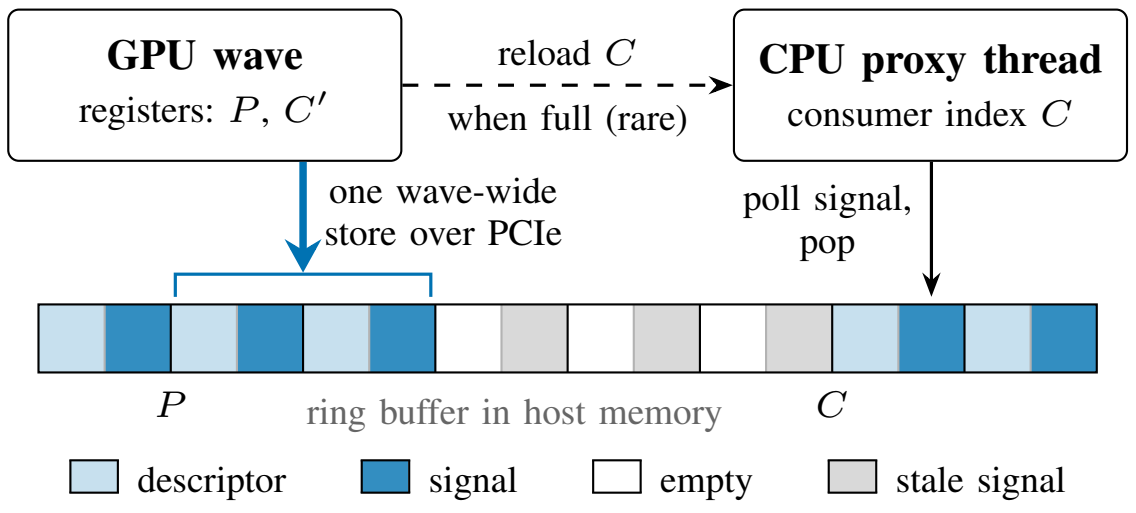
\begin{figure}[t]
\centering
\begin{tikzpicture}[
    font=\footnotesize,
    every node/.style={align=center},
    agent/.style={draw, rounded corners=2pt, minimum width=26mm, minimum height=10mm, inner sep=2pt},
    sub/.style={font=\scriptsize},
    acc/.style={-{Stealth[length=1.6mm]}, semithick},
    rare/.style={-{Stealth[length=1.6mm]}, semithick, dashed},
]
\definecolor{sysblue}{HTML}{0072B2}
\colorlet{pay}{sysblue!22}   %
\colorlet{sig}{sysblue!80}   %
\colorlet{old}{gray!30}       %
\def\cw{0.875}                  %
\def\ch{0.45}                 %
\def\ry{-2.0}                 %
\node[agent] (gpu) at (1.1, -0.55)
  {\textbf{GPU wave}\\[1pt] \scriptsize registers: $P$, $C'$};
\node[agent] (cpu) at (5.9, -0.55)
  {\textbf{CPU proxy thread}\\[1pt] \scriptsize consumer index $C$};
\draw[rare] (gpu.east) -- (cpu.west)
  node[midway, above, sub] {reload $C$} node[midway, below, sub] {when full (rare)};
\foreach \i in {0,...,7} {
  \pgfmathsetmacro{\x}{\i*\cw}
  \pgfmathtruncatemacro{\pub}{(\i<=2 || \i>=6) ? 1 : 0}
  \ifnum\pub=1
    \fill[pay] (\x, \ry) rectangle ++(\cw/2, -\ch);
    \fill[sig] (\x+\cw/2, \ry) rectangle ++(\cw/2, -\ch);
  \else
    \fill[old] (\x+\cw/2, \ry) rectangle ++(\cw/2, -\ch);
  \fi
}
\foreach \i in {0,...,7}
  \draw[gray!60] (\i*\cw+\cw/2, \ry) -- ++(0, -\ch);
\foreach \i in {1,...,7}
  \draw (\i*\cw, \ry) -- ++(0, -\ch);
\draw (0, \ry) rectangle (8*\cw, \ry-\ch);
\node[sub, text=black!60, anchor=north] at (3.15, \ry-\ch-0.05) {ring buffer in host memory};
\newcommand{\swatch}[1]{{\setlength{\fboxsep}{0pt}\setlength{\fboxrule}{0.4pt}\fcolorbox{black}{#1}{\rule{0pt}{2mm}\hspace{3mm}}}}
\node[sub, anchor=north] at (4*\cw, \ry-\ch-0.45)
  {\swatch{pay}\enspace descriptor\qquad\swatch{sig}\enspace signal\qquad\swatch{white}\enspace empty\qquad\swatch{old}\enspace stale signal};
\draw[sysblue, semithick] (1*\cw+0.03, \ry+0.05) -- (1*\cw+0.03, \ry+0.2) -- (3*\cw-0.03, \ry+0.2) -- (3*\cw-0.03, \ry+0.05);
\draw[-{Stealth[length=2.2mm, width=2.4mm]}, line width=1.4pt, sysblue]
  (gpu.south -| 2*\cw, 0) -- (2*\cw, \ry+0.21)
  node[pos=0.45, right, sub, text=black] {one wave-wide\\store over PCIe};
\node[sub, anchor=north] at (1*\cw, \ry-\ch) {$P$};
\draw[acc] (cpu.south -| 6.75*\cw, 0) -- (6.75*\cw, \ry+0.05)
  node[pos=0.45, left, sub] {poll signal,\\pop};
\node[sub, anchor=north] at (6*\cw, \ry-\ch) {$C$};
\end{tikzpicture}
\caption{\sys's SPSC device-to-host queue. $P$ is the producer index, $C$ the consumer index, and $C'$ the wave's cached copy of $C$. Each 8-byte slot pairs 4 bytes of a descriptor with a 4-byte signal written in the same store, so those bytes are valid once their signal is. Blue slots are published but not yet popped.}
\label{fig:sys-queue-ring}
\end{figure}

%% file: sections/p2p.tex
\section{KIWI-P2P: SHMEM-Like Device-Initiated Communication}
\label{sec:sys-p2p}

\sysptp combines \sys with the Lightweight Communication Interface (LCI)~\cite{lci2025} to provide a SHMEM-like device-initiated communication library. It serves two purposes: it shows that \sys can be composed with an existing host communication library to expose new device-initiated operations, and it provides a reverse-offload implementation to compare against state-of-the-art GDA implementations.

LCI is a host communication library designed for multithreaded HPC applications~\cite{lci2025}. It provides portable point-to-point and collective operations over libibverbs~\cite{rdmacore2026} and libfabric~\cite{libfabric2015}, and gives users explicit control over network resources, memory registration, and progress. In particular, an LCI \emph{device} bundles an independent set of network resources, such as ibverbs queue pairs, so threads that use different devices do not interfere. \sysptp uses only LCI's public interface, with a \sys callback translating each descriptor into LCI operations. Any other host library, such as libibverbs or libfabric, would also work, but a high-level library like LCI lets developers build device-initiated operations without low-level RDMA programming complexity.

\sysptp drives communication with multiple host proxy threads. It pairs each thread with its own LCI device: a proxy thread runs the callbacks of its assigned \sys endpoints and progresses that device. With \sys and LCI, a new device-initiated operation needs only three pieces: a descriptor, a device wrapper that invokes a callback with it, and a host callback that translates it into LCI operations. Listing~\ref{lst:sys-p2p-post-put} shows these pieces for \texttt{post\_put\_signal}. Its callback looks up the destination's symmetric-heap registration and posts the payload followed by a one-byte readiness flag on the same LCI device. LCI guarantees that the target sees the updated flag only after the payload. Everything else, including bootstrapping, low-level RDMA programming, and GPU-to-CPU handoff, is reused from \sys and LCI.

\begin{listing}[t]
\centering
\begin{minipage}{0.98\columnwidth}
\scriptsize
\begin{lstlisting}[style=syscode]
struct PutSignalDesc { src, size, rank, heap_id,
                 dst_off, flag_off; };

// GPU: called by all lanes of a wave.
__device__ void post_put_signal(Handle& h,
                                PutSignalDesc d) {
  while (!h.invoke(PUT_CB, d)) {}
}

// CPU: the callback registered as PUT_CB.
void put_signal_callback(PutSignalDesc d) {
  rkey = lookup(d.heap_id, d.rank);
  lci::post_put(d.rank, d.src, d.size, d.dst_off,
                rkey);
  lci::post_put(d.rank, /*flag=*/1, d.flag_off, rkey);
}
\end{lstlisting}
\end{minipage}
\caption{The \sysptp put operation: a descriptor, a device wrapper, and a host callback.}
\label{lst:sys-p2p-post-put}
\end{listing}

\paragraph{Reverse-offload overhead.}
To measure what reverse offload adds, we compare \sysptp with two host-initiated baselines, \emph{flag + LCI} and \emph{flag + ibverbs}. In all three paths, one rank puts a batch of 8-byte values into its peer's GPU memory, a GPU kernel on the peer polls until the whole batch has arrived, and the peer then puts one acknowledgment back. In \sysptp, the GPU kernels issue every put through \sys, and the proxy callback posts it with an LCI put. In the baselines, the GPU instead notifies a host thread through a pinned host flag once a batch or acknowledgment arrives, and the host thread then posts the next puts itself. This flag-based handoff resembles the one fabric-lib uses~\cite{fabriclib2026}. A batch of one measures ping-pong latency, and a batch of 128 measures single-endpoint throughput. Each rank uses one host thread and one queue pair.

Reverse offload costs \sysptp little (Table~\ref{tab:sys-p2p-overhead}). Compared with flag + LCI, \sysptp has the same latency (5.23 versus 5.25~\(\mu\mathrm{s}\)) and loses only 3\% of throughput (2.88 versus 2.97~Mops/s). The GPU hands off descriptors faster than the proxy can post them, so both paths are limited by the same LCI post. Flag + ibverbs reaches 3.40~Mops/s because it avoids LCI's per-put cost, which is independent of reverse offload. Section~\ref{sec:evaluation} compares \sysptp with GDA libraries.

\begin{table}[t]
  \centering
  \caption{Eight-byte single-PUT latency and throughput between two nodes. All paths use GPU buffers and receiver-GPU polling. In the baselines, the GPU notifies a host thread through a pinned host flag, and the host thread posts the puts.}
  \label{tab:sys-p2p-overhead}
  \small
  \setlength{\tabcolsep}{4pt}
  \begin{tabular}{@{}lrr@{}}
    \toprule
    Path & Latency (\(\mu\mathrm{s}\)) & Throughput (Mops/s) \\
    \midrule
    \sysptp (\sys + LCI) & 5.23 & 2.88 \\
    Flag + LCI & 5.25 & 2.97 \\
    Flag + ibverbs & 5.24 & 3.40 \\
    \bottomrule
  \end{tabular}
\end{table}

%% file: sections/ep.tex
\section{KIWI-EP: Expert-Parallel Communication with Workload-Defined Operations}
\label{sec:sys-ep}

Mixture-of-experts (MoE) architectures are now prevalent in frontier large language models, and their communication can become a major bottleneck in MoE serving~\cite{deepep2025,swiftep2026}. MoE layers perform two irregular all-to-all operations: \emph{dispatch} routes each token to its selected experts, and \emph{combine} returns the expert outputs to the token's source and reduces them. This bottleneck becomes particularly important for agentic workloads, which increasingly demand low-latency serving at low to medium concurrency.

\sysptp shows that \sys composes easily with an existing host communication library and delivers competitive performance. Its operations, however, are generic primitives such as put, which GPU kernels must compose into higher-level protocols. This section presents \sysep, a new set of low-latency multi-node MoE dispatch and combine kernels built on \sys and LCI, which go further and offload parts of the EP protocol itself to the CPU. \sysep defines three EP-specific device-initiated operations, a multicast put, a count all-to-all, and a barrier, each implemented as a \sys callback. It is up to $2.3\times$ faster than state-of-the-art EP communication kernels on AMD GPUs.

\subsection{Dispatch and Combine Kernels}

Listings~\ref{lst:sys-ep-dispatch} and~\ref{lst:sys-ep-combine} summarize the \sysep dispatch and combine kernels. Dispatch first packs tokens into network-registered staging buffers, invokes one multicast put per token, and unpacks the received messages into expert input rows. Before packing begins, a prescan computes the per-destination send counts and posts a nonblocking all-to-all; after packing, the kernel waits for the exchanged counts to determine how many messages to receive. Combine reverses this process by packing each expert output row, posting a put to its source rank, and accumulating the returned rows into the corresponding tokens. It also posts a nonblocking barrier to prevent the next dispatch from overwriting the staging buffers before all ranks have finished using them.

\begin{listing}[t]
\centering
\begin{minipage}{0.98\columnwidth}
\scriptsize
\begin{lstlisting}[style=syscode]
// Prescan and exchange counts.
workgroup 0:
  send_counts = count_unique_dests(topk_indices)
  count_ticket = post_all2all(send_counts)

// Pack and send token rows.
parallel_for token in tokens [round-robin over waves]:
  dests = deduplicate(topk_indices[token])
  payload = pack_to_staging(token, route_metadata)
  post_put_multicast(remote_peers(dests), payload)
  copy_to_self_and_local_peers(dests, payload)

// Determine the receive layout, then unpack.
workgroup 0:
  wait_all2all(count_ticket, recv_counts)
  build_receive_map(recv_counts)
grid_barrier()

parallel_for row in recv_rows [round-robin over waves]:
  wait_if_remote(readiness_flag[row])
  unpack_to_expert_input(row)
\end{lstlisting}
\end{minipage}
\caption{Simplified \sysep dispatch kernel.}
\label{lst:sys-ep-dispatch}
\end{listing}

\begin{listing}[t]
\centering
\begin{minipage}{0.98\columnwidth}
\scriptsize
\begin{lstlisting}[style=syscode]
// Pack and return expert output rows.
parallel_for row in outputs [contiguous over waves]:
  payload = pack_to_staging(row, route_metadata)
  post_put(source_rank(row), payload,
              combine_readiness_flag(row))

grid_barrier()
workgroup 0, wave 0:
  barrier_ticket = post_barrier()

// Wait for returned rows and reconstruct tokens.
parallel_for token in tokens [round-robin over waves]:
  wait(combine_readiness_flags[token])
  accumulate_into_output(token)

grid_barrier()
workgroup 0:
  reset_combine_readiness_flags()
  wave 0: wait_barrier(barrier_ticket)
\end{lstlisting}
\end{minipage}
\caption{Simplified \sysep combine kernel.}
\label{lst:sys-ep-combine}
\end{listing}

The EP algorithm stays straightforward because \sys lets \sysep express its communication as three EP-specific operations: a multicast put, a nonblocking count all-to-all, and a nonblocking barrier. Each is a single \sys invocation from the calling wave, and its host callback performs the protocol work that would otherwise run on GPU threads. The two nonblocking operations are split into a post and a wait, so the kernel can overlap their progress with payload work.

The dispatch kernel sends each token to the ranks selected by its top-$k$ routing with one multicast put. The wave builds the descriptor cooperatively in LDS, listing up to eight destination ranks with their remote offsets and readiness slots, and submits it with a single invocation. One host callback then posts the payload and readiness writes for every destination.

The dispatch kernel uses the nonblocking count all-to-all to exchange per-destination token counts. The exchange is hierarchical, using RDMA between nodes and IPC loads within a node. In the post phase, the GPU packs the counts for each remote node into one contiguous row and submits a single \sys invocation, whose host callback issues one LCI put per remote node rather than one per destination GPU. In the wait phase, GPU lanes poll the ingress buffers of their local peers through IPC loads and copy out the counts. Each count travels in one 64-bit word together with a 32-bit epoch, so a lane detects its arrival from the epoch without a separate signal write.

The barrier currently reuses the hierarchical exchange underlying the count all-to-all. For larger barrier domains, its cross-node component could adopt a more scalable algorithm, such as the dissemination barrier~\cite{disseminationbarrier1988}. Such a protocol would require substantial device-side machinery under GDA but can be implemented more directly with programmable host proxy threads.

\subsection{Dynamic KIWI Endpoint Polling}

In real serving workloads, continuous batching changes the active token count of each dispatch or combine invocation from step to step. \sysep launches a different number of workgroups (thread blocks) accordingly. Because each wave owns one \sys endpoint, the number of active endpoints changes as well. vLLM initializes the \sys backend only once, however, so its persistent proxy threads must adapt their polling ranges without being restarted.

To track the active range, wave~0 of workgroup~0 in every dispatch and combine kernel invokes an active-endpoint-count callback with the launch's endpoint count. Issuing the callback from inside the kernel keeps it valid under HIP-graph replay. The callback publishes the count in a global variable. Each proxy sweeps its endpoints in increasing order, rereads this variable once per sweep, and stops at the active boundary. A proxy grows its range immediately; on a shrink, it keeps polling the retiring endpoints until they drain.

\input{figures/ep_polling}

The proxies must also share the polling work evenly, which is complicated by two factors. First, the active endpoint count changes from step to step. Second, requests do not arrive on contiguous endpoints: when several waves cooperate on one token, only the first wave posts the put, so with one token per workgroup only wave~0 of each workgroup submits requests, giving a strided pattern.

Figure~\ref{fig:sys-ep-polling} compares three ways to assign endpoints to $P$ proxy threads. \emph{Contiguous} assignment gives each proxy a fixed range of endpoints. Because the active count changes, the ranges must cover all provisioned endpoints, so when only a few workgroups are active, all of their endpoints fall on one proxy. \emph{Round-robin} assignment gives endpoint $q$ to proxy $q \bmod P$ and balances the active endpoints, but not strided requests. With 16 waves per workgroup and $P = 4$, for example, the wave-0 endpoints $0, 16, 32, \ldots$ all map to proxy~0. \sysep instead uses \emph{skewed round-robin}, in which the owner of endpoint $q$ is the sum of the digits of $q$ written in base $P$, modulo $P$. With $P = 4$, endpoint $16 = 100_4$ thus belongs to proxy~1. Every aligned group of $P$ endpoints covers all $P$ proxies, which balances any active prefix. When the number of waves per workgroup and $P$ are both powers of two, as in our configuration (16 waves and $P = 4$), the same holds one level up: the wave-0 endpoints of every aligned group of $P$ workgroups land on different proxies, which balances the strided requests.

%% file: figures/ep_polling.tex
\begin{figure}[t]
\centering
\definecolor{proxy0}{HTML}{2A78D6}
\definecolor{proxy1}{HTML}{EB6834}
\definecolor{proxy2}{HTML}{1BAF7A}
\definecolor{proxy3}{HTML}{4A3AA7}
\begin{tikzpicture}[x=9.2pt, y=-6.6pt, font=\scriptsize]
  \node[anchor=east, font=\tiny, text=black!70] at (-0.15, -0.7) {wave};
  \foreach \v in {0,...,15}
    \node[font=\tiny, text=black!70] at (\v+0.5, -0.7) {\v};
  \foreach \name/\mode [count=\s from 0] in {Contiguous/0, Round-robin/1, {Skewed\\round-robin}/2} {
    \pgfmathsetmacro{\yoff}{\s*5.3}
    \node[anchor=east, align=right] at (-2.0, \yoff+2) {\name};
    \foreach \w in {0,...,3} {
      \node[anchor=east, font=\tiny, text=black!70] at (-0.15, \yoff+\w+0.5) {WG\w};
      \foreach \v in {0,...,15} {
        \pgfmathtruncatemacro{\q}{16*\w+\v}
        \ifcase\mode
          \def\own{0}%
        \or
          \pgfmathtruncatemacro{\own}{mod(\q,4)}%
        \or
          \pgfmathtruncatemacro{\own}{mod(mod(\q,4)+mod(floor(\q/4),4)+floor(\q/16),4)}%
        \fi
        \ifnum\own=0 \def\ink{white}\fi
        \ifnum\own=1 \def\ink{black}\fi
        \ifnum\own=2 \def\ink{black}\fi
        \ifnum\own=3 \def\ink{white}\fi
        \fill[proxy\own, draw=white, line width=0.5pt] (\v,\yoff+\w) rectangle +(1,1);
        \node[font=\fontsize{4.5}{5}\selectfont, text=\ink, inner sep=0pt]
          at (\v+0.5, \yoff+\w+0.5) {\own};
      }
      \draw[black, line width=0.9pt] (0,\yoff+\w) rectangle +(1,1);
    }
  }
  \foreach \p in {0,...,3} {
    \fill[proxy\p] (\p*4, 16.0) rectangle +(1,1);
    \node[anchor=west, inner sep=1pt] at (\p*4+1.1, 16.5) {Proxy~\p};
  }
\end{tikzpicture}
\caption{Endpoint-to-proxy assignment with four proxies and 4 of 32 provisioned workgroups active. Each cell is an endpoint. Outlined cells are wave-0 endpoints, which carry the requests. Only skewed round-robin spreads both the active and the outlined endpoints across all proxies.}
\label{fig:sys-ep-polling}
\end{figure}
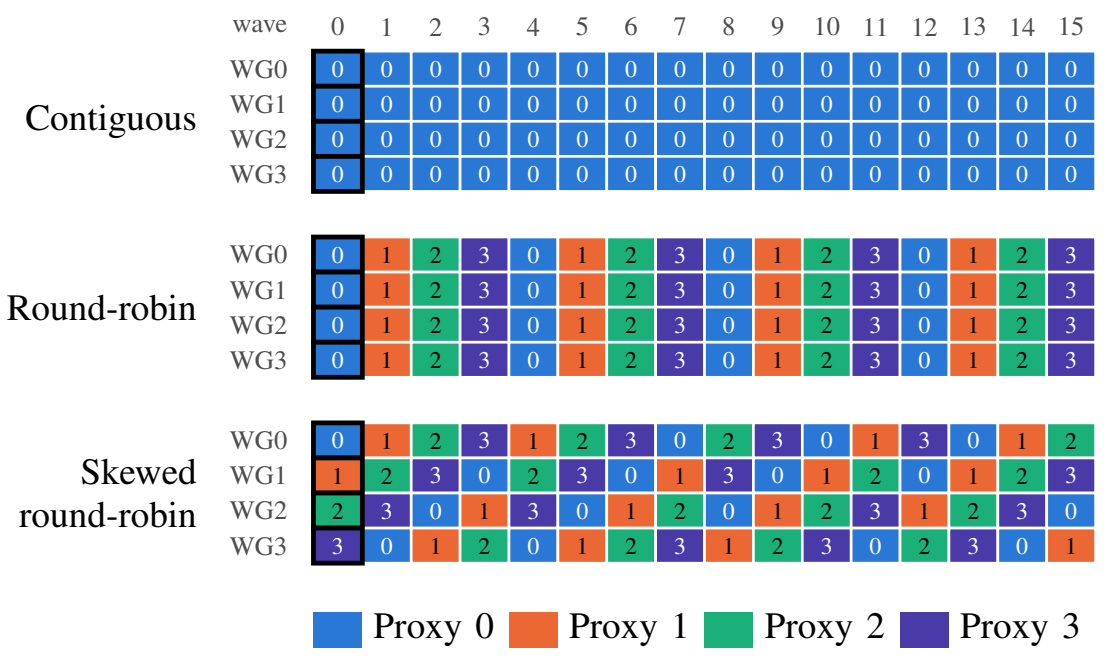

%% file: sections/evaluation.tex
\section{Evaluation}
\label{sec:evaluation}

Our evaluation answers three questions:
\begin{enumerate}[leftmargin=*,nosep,label=\textbf{Q\arabic*.}]
  \item How cheap is \sys's GPU-to-CPU handoff, and how does it compare with the existing mechanisms in other libraries as concurrency increases?
  \item How does \sysptp compare with state-of-the-art device-initiated communication libraries on basic point-to-point primitives?
  \item How does \sysep compare with state-of-the-art expert-parallel communication libraries, both at the kernel level and end to end, and how much of its benefit comes from each optimization?
\end{enumerate}

\subsection{Experimental Setup}

\input{tables/setup}

We run all experiments on an AMD Instinct\texttrademark{} MI300X cluster with a RoCE network; Table~\ref{tab:setup} lists its hardware configuration. Each node has two 56-core Intel Xeon Platinum 8480+ CPUs, eight MI300X GPUs, and eight 400~Gb/s NVIDIA ConnectX-7 NICs. Each GPU shares a PCIe switch with one NIC, so GPUDirect RDMA traffic bypasses the CPU root complex. We pin each process and its proxy threads to CPU cores in the NUMA node of its GPU, and each process uses the NIC that shares its GPU's PCIe switch. Each data point is the median of at least three runs.

\subsection{GPU-to-CPU Handoff}
\input{tables/queue_baselines}
\begin{figure*}[t]
\centering
\begin{minipage}[t]{0.48\textwidth}
\centering
  \includegraphics[width=\linewidth]{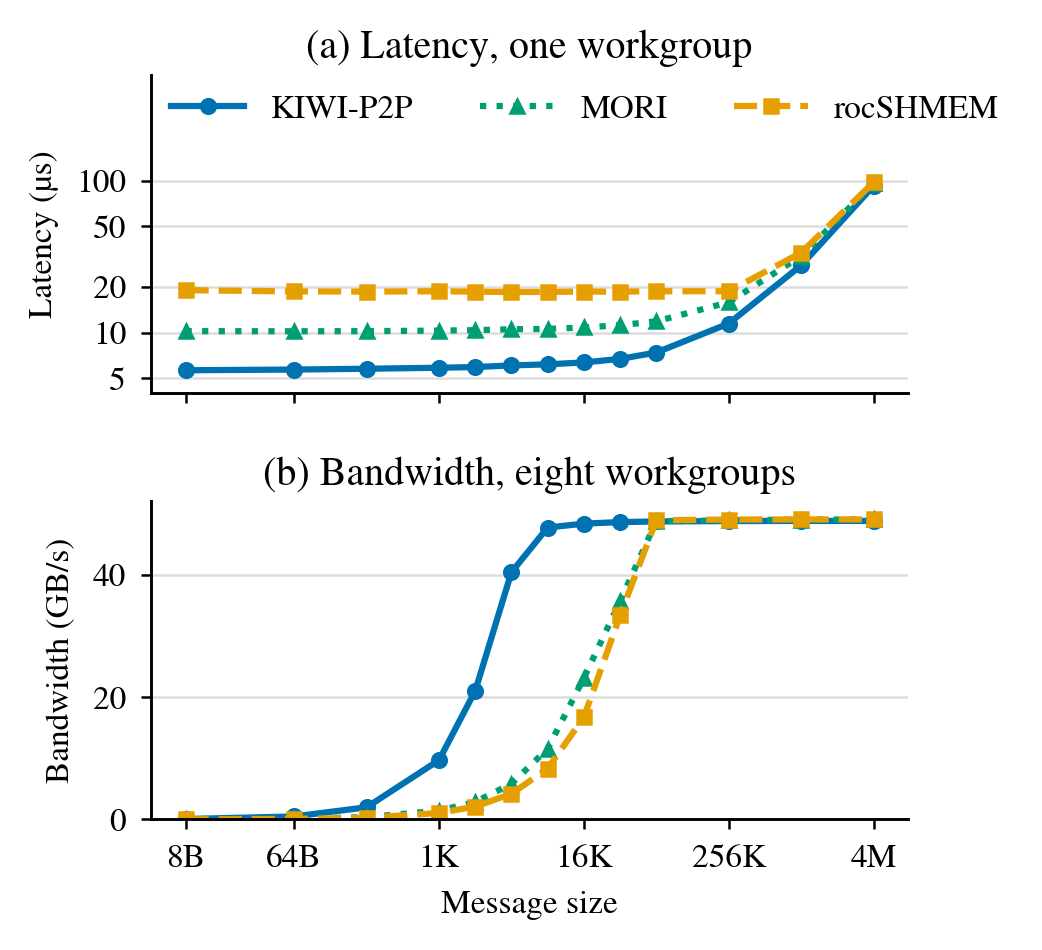}
  \caption{Point-to-point (a) one-way latency with one workgroup and (b) bandwidth with eight workgroups. MORI and rocSHMEM use GDA. Two nodes, one rank per node.}
  \label{fig:p2p}
\end{minipage}\hfill
\begin{minipage}[t]{0.48\textwidth}
\centering
  \includegraphics[width=\linewidth]{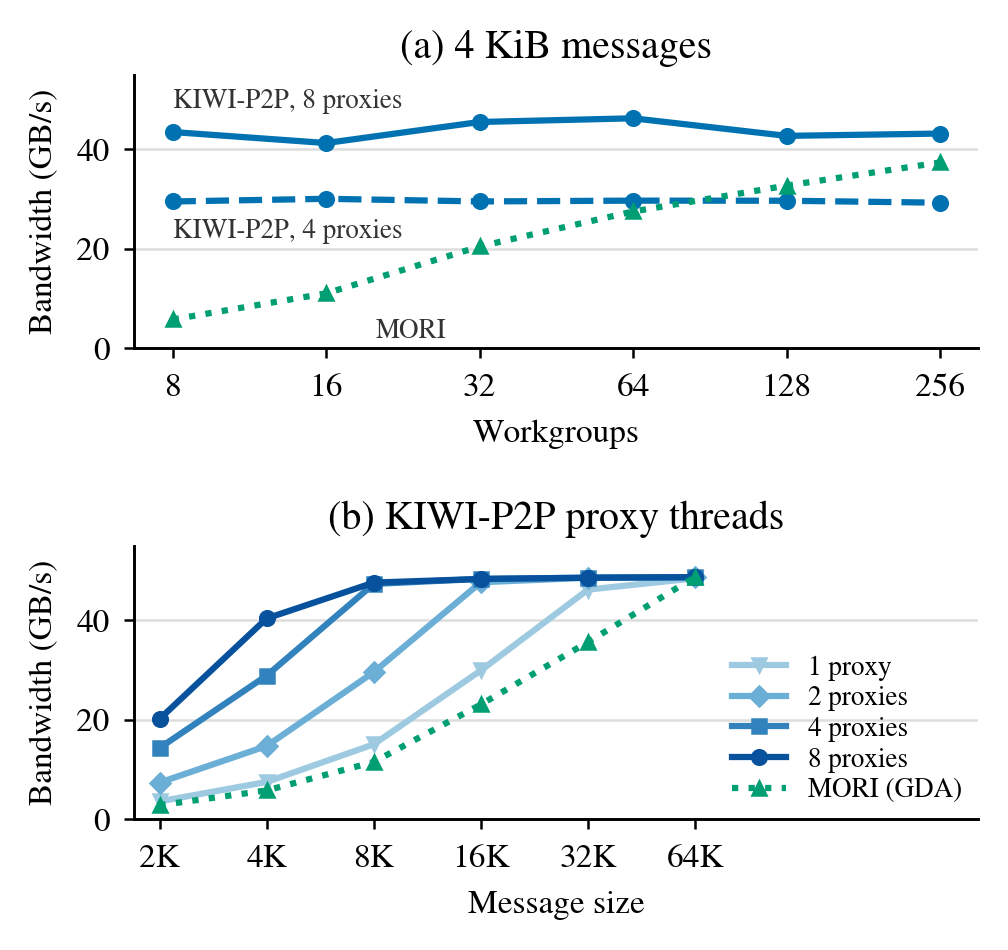}
  \caption{Point-to-point bandwidth scaling. (a) Bandwidth versus the number of producer workgroups for 4~KiB messages; MORI uses its best number of queue pairs. (b) \sysptp bandwidth with eight workgroups and one to eight proxy threads, with MORI at eight workgroups for reference. Two nodes, one rank per node.}
  \label{fig:p2p-scaling}
\end{minipage}
\end{figure*}

We first ask whether \sys's handoff is as cheap as Section~\ref{sec:sys-queue} suggests, and how it compares with the GPU-to-CPU queues of six existing systems (Q1).  We evaluate all queues in a common harness in which only the queue's own handoff protocol differs. NVSHMEM does not run on AMD GPUs, and rocSHMEM's reverse-offload queue is internal to the library, so we re-implement their publication protocols following the upstream source. All the other queues run their unmodified upstream device code. RCCL GIN's proxy queue is a port of NCCL GIN's, so its results also represent NCCL GIN; we enable its faster GDRCopy path, which keeps the consumer index in GPU memory rather than host memory. To measure the round trip, a single wave pushes one message, a proxy thread pops it and writes an acknowledgement flag, and the wave waits for the flag before pushing the next message. As a lower bound, a raw-store baseline issues one wave-wide store in place of a queue push.

\begin{figure*}[t]
  \centering
  \includegraphics[width=\textwidth]{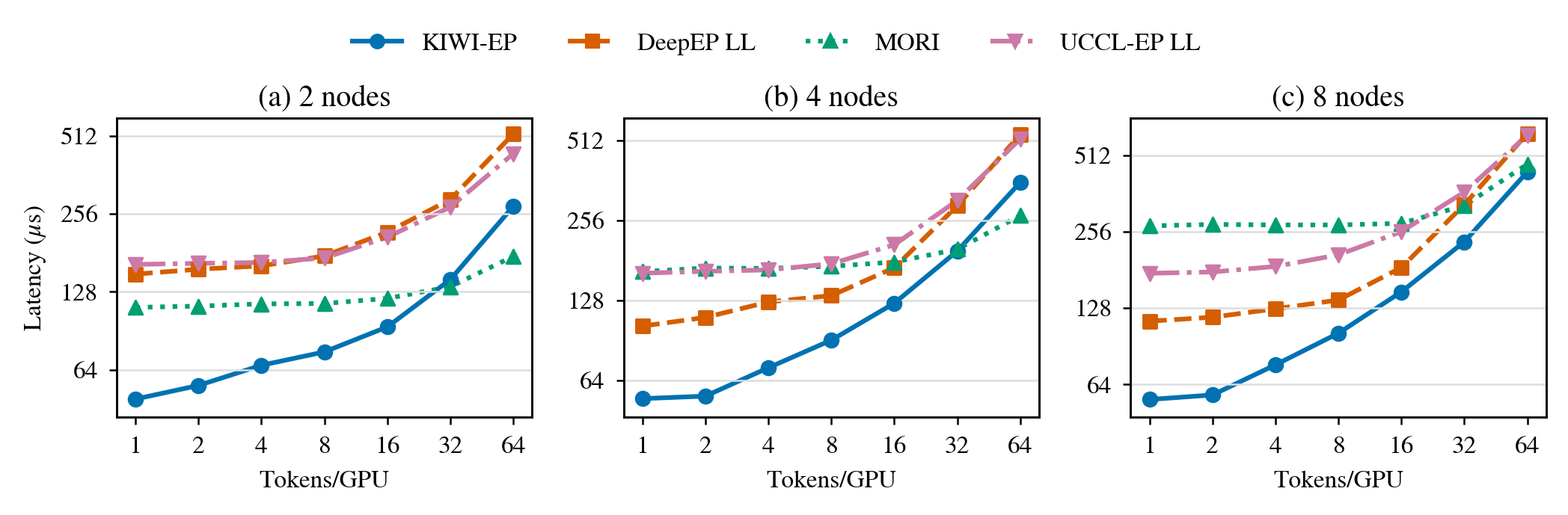}
  \caption{Expert-parallel dispatch-plus-combine latency on two to eight nodes; lower is better.}
  \label{fig:ep}
\end{figure*}

Table~\ref{tab:queue-baselines} shows that the \sys queue completes the round trip in 1.92~\(\mu\mathrm{s}\) at the median, the same as the raw store, so queue management adds almost nothing beyond the store itself; the fastest existing queues take 2.68--2.80~\(\mu\mathrm{s}\). The gap widens as more waves produce concurrently. Because each wave owns its queue and publishes without atomics, the \sys queue's push cost stays at 0.12~\(\mu\mathrm{s}\) as the number of producer waves grows from 1 to 1,024, whereas that of every existing queue grows by 4.6--16$\times$. The same trend shows in throughput: with 256 waves, the \sys queue sustains 3.3$\times$ the throughput of the best existing queue. The host side is equally cheap: polling an empty \sys queue costs 0.87~ns because the poll hits in the CPU L1 cache, so one proxy thread can monitor many endpoints.

Programmability adds almost no cost on top. With callbacks, the median round trip (1.92~\(\mu\mathrm{s}\)) and the single-wave push cost (0.12~\(\mu\mathrm{s}\)) are unchanged; at 1,024 waves, the push costs 0.16~\(\mu\mathrm{s}\). At 256 waves, throughput drops to 62\% of the bare queue's because the proxy threads must run a callback for every message. Kernels written in FlyDSL and Triton (Section~\ref{sec:sys-frontends}) match HIP C++'s round trip to within 2\% and reach 78--96\% of its single-endpoint throughput.

\subsection{Point-to-Point Communication}

We next ask whether reverse offload can match GDA on basic point-to-point primitives (Q2). We compare \sysptp with the GDA paths of MORI and rocSHMEM~\cite{mori2025,rocshmem2026} using a common benchmark with matched kernels. In the latency test, one workgroup on each of two ranks exchanges puts in a ping-pong, and the receiver observes each payload before responding. In the bandwidth test, eight workgroups stream puts to the peer, and each system uses its own per-workgroup resources: \sysptp assigns one \sys endpoint to each workgroup and serves them with eight proxy threads, while MORI and rocSHMEM give each workgroup its own queue pair or context.

Figure~\ref{fig:p2p} shows that \sysptp has the lowest latency at every message size. A small put takes 5.64~\(\mu\mathrm{s}\), 1.8$\times$ and 3.4$\times$ lower than MORI and rocSHMEM, and the systems converge only once transfer time dominates. The difference lies in the control path. A \sysptp put costs the GPU a single queue push, and the CPU proxy builds the work request and rings the doorbell, whereas a GDA thread performs these steps itself through system-scope memory operations. The same effect appears in bandwidth: with eight workgroups, \sysptp nearly saturates the link at 8~KiB, whereas MORI and rocSHMEM need 64~KiB.

The two designs also scale differently (Figure~\ref{fig:p2p-scaling}). Because each GDA producer pays the full control-path cost, MORI needs many workgroups to fill the link with small messages; even with 128 workgroups, its 4~KiB bandwidth stays below that of \sysptp with eight workgroups. \sysptp's bandwidth is instead set by the number of proxy threads (Figure~\ref{fig:p2p-scaling}b). Eight proxy threads per GPU drive a 400~Gb/s NIC near its payload limit from 8~KiB.

\subsection{Expert-Parallel Communication}

We now evaluate whether workload-defined operations pay off in a real communication workload (Q3). We compare \sysep with ROCm DeepEP, MORI-EP, and UCCL-EP~\cite{rocmdeepep2025,moriep2025,uep2026} on two to eight nodes, using a common harness that gives every backend identical inputs. The workload uses DeepSeek-V3's MoE dimensions (hidden width 7168, 256 experts, and top-8 routing) with BF16 dispatch and combine. We report the latency of one dispatch-plus-combine pair on the slowest rank.

Figure~\ref{fig:ep} shows that \sysep has the lowest latency up to 16 tokens per GPU on two nodes, up to 32 on four, and at every token count on eight nodes, with up to 2.3$\times$ lower latency than the best competing backend. At higher token counts, RDMA bandwidth becomes the bottleneck: \sysep sends a token once per destination GPU, whereas MORI sends it once per destination node and forwards it to the destination GPUs within that node. We leave extending \sysep with such hierarchical token routing to future work.

\begin{figure}[t]
  \centering
  \includegraphics[width=\columnwidth]{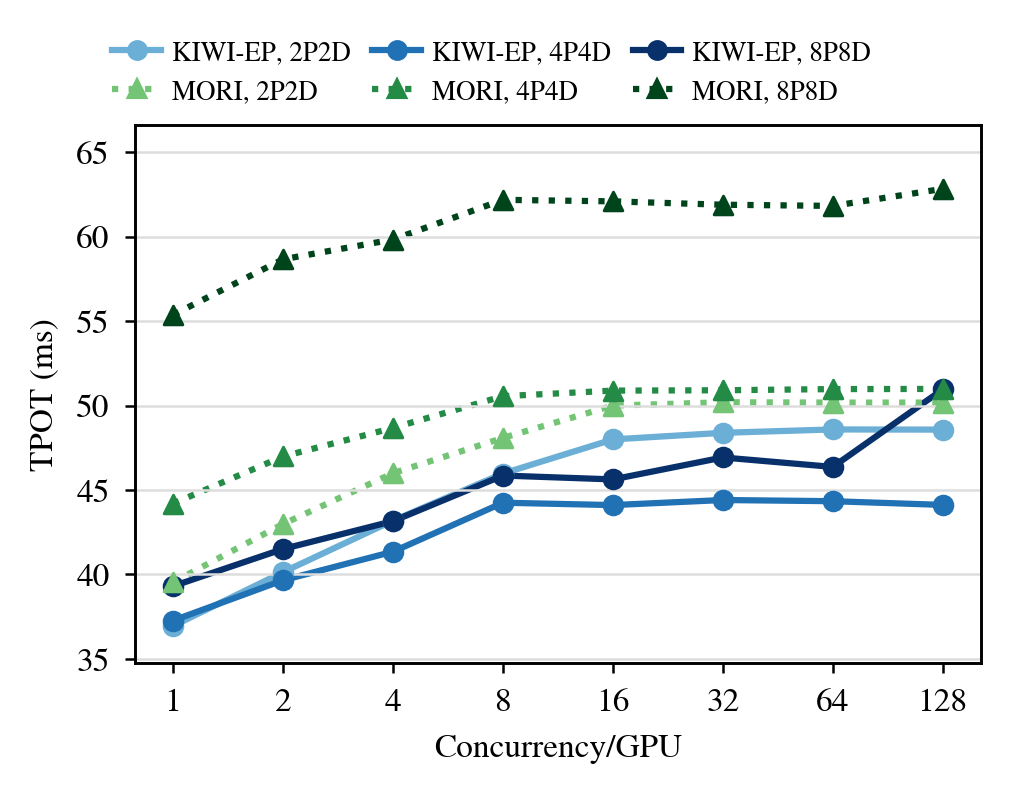}
  \caption{vLLM TPOT for GLM-5.3; lower is better. \(N\)P\(N\)D denotes \(N\) prefill and \(N\) decode nodes. Each node has eight GPUs.}
  \label{fig:vllm}
\end{figure}

\begin{figure}[t]
  \centering
  \includegraphics[width=\columnwidth]{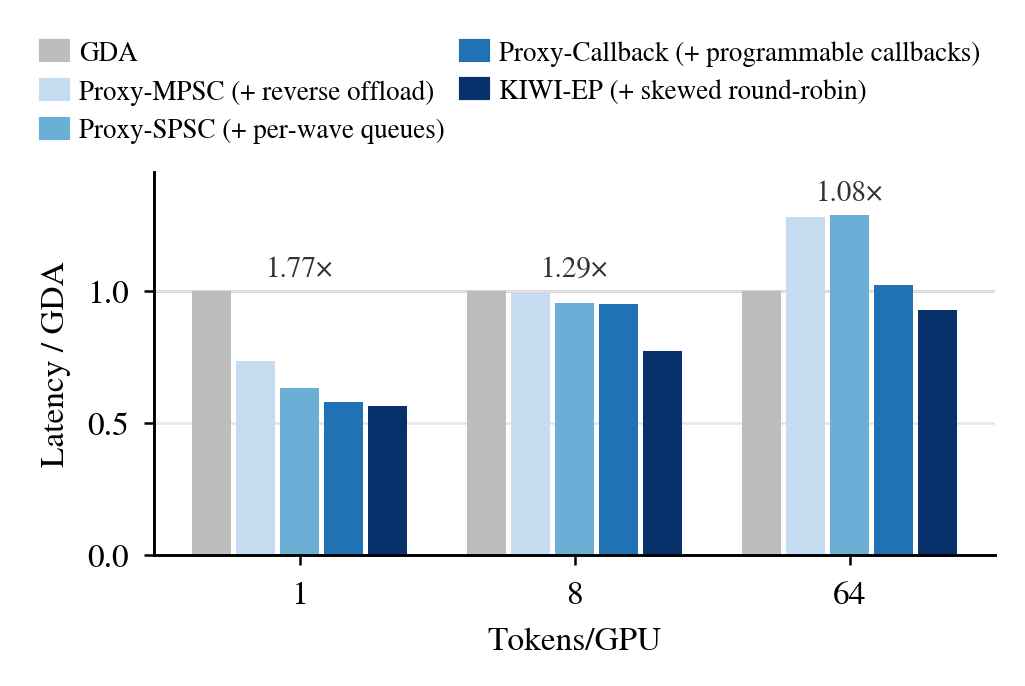}
  \caption{\sysep ablation ladder on eight nodes. Bars give the dispatch-plus-combine latency of each variant normalized to GDA; lower is better. Each variant adds one design choice to the previous one. Labels give the speedup of \sysep over GDA.}
  \label{fig:sys-ep-ablation}
\end{figure}

These kernel-level gains carry over to serving. We integrate \sysep into vLLM~\cite{vllm2023} and serve GLM-5.3~\cite{glm5report2026,glm53modelcard2026}, one of the most widely used open-weight MoE models, in FP8 with prefill-decode (PD) disaggregation. Each deployment has \(N\) prefill and \(N\) decode nodes (\(N\)P\(N\)D) with eight data- and expert-parallel ranks per node, and requests have 8K input and 1K output tokens. Prefill always uses MORI-EP, the recommended EP backend for vLLM on AMD GPUs, and decode uses either MORI-EP or \sysep. Due to PD disaggregation and continuous batching, the number of requests that decode runs at a time is lower than the configured concurrency, so the TPOT gains do not map strictly onto the kernel-level gains. Figure~\ref{fig:vllm} shows that \sysep lowers TPOT over MORI at every scale and concurrency we measured, by up to 29\% on eight nodes. We also run GSM8K~\cite{cobbe2021gsm8k} and needle-in-a-haystack~\cite{kamradt2023niah} as accuracy tests and verify that \sysep does not affect accuracy.

To understand where these gains come from, we build five variants of \sysep, each adding one design choice to the previous one (Figure~\ref{fig:sys-ep-ablation}). \emph{GDA} posts every put directly from the GPU, using an mlx5 implementation modeled on MORI. \emph{Proxy-MPSC} switches to reverse offload using only the put primitives, and all producer waves share one MPSC \sys queue per proxy thread. \emph{Proxy-SPSC} gives each wave its own SPSC queue, and \emph{Proxy-Callback} replaces generic puts with \sysep's workload-defined callbacks. These variants assign endpoints to proxy threads round-robin; \sysep finally adds the skewed round-robin mapping (Figure~\ref{fig:sys-ep-polling}).

Together, these choices make \sysep 1.77$\times$ faster than GDA at one token per GPU, 1.29$\times$ at eight, and 1.08$\times$ at 64. Reverse offload helps most at small batch sizes: at one token per GPU, Proxy-MPSC already has 27\% lower latency than GDA, because a GPU producer only pushes a descriptor instead of building an RDMA work request and ringing the doorbell itself, and per-wave SPSC queues cut latency by another 14\%. With large batches, however, round-robin assignment creates imbalance between proxy threads, and the two proxy variants have 28\% and 29\% higher latency than GDA at 64 tokens. Programmable callbacks shrink the proxy work by fusing each payload with its signal and coalescing multicasts and collectives, cutting latency by 21\%; skewed round-robin then spreads the requests over all proxy threads, cutting latency by another 19\% at eight tokens and 9\% at 64. We also measured the gain from programmable callbacks across per-token payload sizes (not shown): the gain grows as payloads shrink, because the per-message control work that callbacks remove does not shrink with the payload. We therefore expect callbacks to matter more as recent models dispatch smaller per-token payloads~\cite{nvfp42025,latentmoe2026}.

%% file: tables/setup.tex
\begin{table}[t]
  \centering
  \caption{Hardware configuration of the evaluation cluster.}
  \label{tab:setup}
  \scriptsize
  \setlength{\tabcolsep}{3pt}
  \begin{tabular}{@{}lp{0.76\columnwidth}@{}}
    \toprule
    \multicolumn{2}{@{}l}{\textit{Hardware (per node)}} \\
    CPU & 2$\times$ Intel Xeon Platinum 8480+ (56 cores each), 2~TiB memory \\
    GPU & 8$\times$ AMD Instinct MI300X (192~GB HBM3) \\
    NIC & 8$\times$ NVIDIA ConnectX-7, 400~Gb/s RoCE \\
    PCIe & 8$\times$ Broadcom PEX890xx switch, Gen5 x16 (64~GB/s) \\
    \bottomrule
  \end{tabular}
\end{table}

%% file: tables/queue_baselines.tex
\begin{table}[t]
  \centering
  \caption{GPU-to-CPU queue comparison on one MI300X. $W$ is the number of producer waves. RTT is the unloaded GPU--CPU--GPU round trip, push is the p50 device-side cost of one push, and idle poll is the host cost of polling one empty queue. \emph{Own} gives each wave its own queue; \emph{shared} puts all waves on one queue. $^\dagger$Re-implemented on AMD GPUs.}
  \label{tab:queue-baselines}
  \scriptsize
  \setlength{\tabcolsep}{1.5pt}
  \begin{tabular}{@{}lrrrrrrrrr@{}}
    \toprule
    & \multicolumn{2}{c}{RTT (\(\mu\mathrm{s}\))} & \multicolumn{3}{c}{Push (\(\mu\mathrm{s}\))} & \multicolumn{3}{c}{Mops/s} & Idle \\
    \cmidrule(lr){2-3}\cmidrule(lr){4-6}\cmidrule(lr){7-9}
    & & & & \multicolumn{2}{c}{$W$=1{,}024} & & \multicolumn{2}{c}{$W$=256} & poll \\
    \cmidrule(lr){5-6}\cmidrule(lr){8-9}
    Queue & p50 & p99 & $W$=1 & own & shared & $W$=1 & own & shared & (ns) \\
    \midrule
    Raw store (floor) & 1.92 & 1.96 & 0.12 & 0.12 & --- & --- & --- & --- & --- \\
    \sys queue & \textbf{1.92} & \textbf{1.98} & \textbf{0.12} & \textbf{0.12} & --- & \textbf{10.8} & \textbf{392.0} & --- & \textbf{0.87} \\
    RCCL GIN & 3.86 & 4.46 & 1.32 & 6.04 & 7.72 & 1.0 & 39.1 & 4.8 & 20.9 \\
    NVSHMEM$^\dagger$ & 2.68 & 2.84 & 0.92 & 14.92 & 11.84 & 1.0 & 55.7 & 38.5 & 1.67 \\
    MSCCL++ & 2.72 & 2.76 & 0.92 & 7.32 & 7.76 & 1.0 & 119.4 & \textbf{38.9} & 2.44 \\
    UCCL-EP & 2.80 & 2.84 & 1.00 & 5.56 & \textbf{6.12} & 0.9 & 111.3 & 34.5 & 3.13 \\
    LLVM libc RPC & 7.92 & 8.84 & 6.04 & 38.72 & 2{,}032 & 0.2 & 16.4 & 1.5 & 1{,}379 \\
    rocSHMEM$^\dagger$ & 9.28 & 9.60 & 3.84 & 55.64 & 14{,}315 & 0.2 & 2.0 & 0.07 & 1{,}356 \\
    \bottomrule
  \end{tabular}
\end{table}

%% file: sections/conclusion.tex
\section{Discussion and Conclusion}
\label{sec:conclusion}

\paragraph{Toward a better device-initiated interface.} \sys is also an experiment in what a device-initiated communication interface should look like. Existing libraries define a fixed set of operations and their semantics for every kernel; \sys instead keeps a minimal core that hands requests from GPU kernels to host callbacks and leaves the operations and their semantics to the application. Which operations recur across workloads, and which deserve a standard interface, remain open questions that \sys makes cheap to explore.

\paragraph{Beyond communication.} The callback interface is not limited to RDMA. Kernels could also use it to drive GPU copy engines, storage I/O, or sparse computation on the CPU, wherever serial control runs more efficiently on the CPU than on the GPU. As AI agents make custom operations easy to write, developers will write more of them, but maintaining them inside vendor libraries remains difficult. This makes an open, programmable layer like \sys even more valuable.

\paragraph{Proxy thread limitations.} \sys dedicates a few CPU cores per GPU to proxy threads (four in our \sysep deployment), which we consider acceptable because CPU core counts continue to grow and the host CPUs of production GPU servers are often underutilized~\cite{mlaas2022,acme2024}. These proxies also run as ordinary, preemptible operating-system threads. We have not observed preemption to hurt performance in our experiments, but dedicated scheduling support in the operating system could make proxy execution more predictable.

\paragraph{Conclusion.} We have shown that reverse offload, given a cheap GPU-to-CPU handoff, outperforms state-of-the-art GDA implementations, and that making it programmable improves performance further. With \sys, a new device-initiated operation is an ordinary host function rather than a library patch, so kernel developers can define operations that fit their workload without being restricted to vendor-provided primitives. In this way, \sys opens up the design space of communication libraries and fused kernels.

%% file: main.bbl
\begin{thebibliography}{10}
\providecommand{\url}[1]{#1}
\csname url@samestyle\endcsname
\providecommand{\newblock}{\relax}
\providecommand{\bibinfo}[2]{#2}
\providecommand{\BIBentrySTDinterwordspacing}{\spaceskip=0pt\relax}
\providecommand{\BIBentryALTinterwordstretchfactor}{4}
\providecommand{\BIBentryALTinterwordspacing}{\spaceskip=\fontdimen2\font plus
\BIBentryALTinterwordstretchfactor\fontdimen3\font minus \fontdimen4\font\relax}
\providecommand{\BIBforeignlanguage}[2]{{%
\expandafter\ifx\csname l@#1\endcsname\relax
\typeout{** WARNING: IEEEtran.bst: No hyphenation pattern has been}%
\typeout{** loaded for the language `#1'. Using the pattern for}%
\typeout{** the default language instead.}%
\else
\language=\csname l@#1\endcsname
\fi
#2}}
\providecommand{\BIBdecl}{\relax}
\BIBdecl

\bibitem{comet2025}
S.~Zhang, N.~Zheng, H.~Lin, Z.~Jiang, W.~Bao, C.~Jiang \emph{et~al.}, ``{COMET: Fine-grained Computation-communication Overlapping for Mixture-of-Experts},'' in \emph{Proceedings of Machine Learning and Systems}, vol.~7, 2025.

\bibitem{fusedcollectives2024}
K.~Punniyamurthy, K.~Hamidouche, and B.~M. Beckmann, ``{Optimizing Distributed ML Communication with Fused Computation-Collective Operations},'' in \emph{SC24: International Conference for High Performance Computing, Networking, Storage and Analysis}, 2024, pp. 1--17.

\bibitem{mpk2026}
X.~Cheng, Z.~Zhang, Y.~Zhou, J.~Ji, J.~Jiang, Z.~Zhao \emph{et~al.}, ``{MPK: A Compiler and Runtime for Mega-Kernelizing Tensor Programs},'' in \emph{20th USENIX Symposium on Operating Systems Design and Implementation}.\hskip 1em plus 0.5em minus 0.4em\relax USENIX Association, 2026.

\bibitem{deepep2025}
C.~Zhao, S.~Zhou, L.~Zhang, C.~Deng, Z.~Xu, Y.~Liu, K.~Yu, J.~Li, and L.~Zhao, ``{DeepEP: An Efficient Expert-Parallel Communication Library},'' \href{https://github.com/deepseek-ai/DeepEP}{GitHub repository}, 2025, accessed August 23, 2026.

\bibitem{nvshmem2020}
C.-H. Hsu, N.~Imam, A.~Langer, S.~Potluri, and C.~J. Newburn, ``{An Initial Assessment of NVSHMEM for High Performance Computing},'' in \emph{2020 IEEE International Parallel and Distributed Processing Symposium Workshops}, 2020, pp. 1--10.

\bibitem{rocshmem2020}
K.~Hamidouche and M.~LeBeane, ``{GPU Initiated OpenSHMEM: Correct and Efficient Intra-Kernel Networking for dGPUs},'' in \emph{Proceedings of the ACM SIGPLAN Symposium on Principles and Practice of Parallel Programming}, 2020, pp. 336--347.

\bibitem{gpudirectasync2022}
P.~Markthub, J.~Dinan, S.~Potluri, and S.~Howell, ``{Improving Network Performance of HPC Systems Using NVIDIA Magnum IO NVSHMEM and GPUDirect Async},'' \href{https://developer.nvidia.com/blog/improving-network-performance-of-hpc-systems-using-nvidia-magnum-io-nvshmem-and-gpudirect-async/}{NVIDIA Technical Blog}, Nov. 2022, accessed August 23, 2026.

\bibitem{deepseekv3technicalreport}
{DeepSeek-AI}, A.~Liu, B.~Feng, B.~Xue, B.~Wang, B.~Wu \emph{et~al.}, ``{DeepSeek-V3 Technical Report},'' \emph{arXiv preprint arXiv:2412.19437}, 2024.

\bibitem{lci2025}
J.~Yan and M.~Snir, ``{LCI: A Lightweight Communication Interface for Efficient Asynchronous Multithreaded Communication},'' in \emph{Proceedings of the International Conference for High Performance Computing, Networking, Storage and Analysis}, 2025, pp. 1043--1059.

\bibitem{mpi50}
{Message Passing Interface Forum}, \emph{{MPI: A Message-Passing Interface Standard, Version 5.0}}, Jun. 2025.

\bibitem{openshmem16}
{OpenSHMEM Specification Committee}, \emph{{OpenSHMEM Application Programming Interface, Version 1.6}}, Nov. 2024.

\bibitem{nccl2026}
{NVIDIA}, ``{NCCL: NVIDIA Collective Communications Library},'' \href{https://github.com/NVIDIA/nccl}{GitHub repository}, 2026, accessed October 6, 2026.

\bibitem{demystifyingnccl2025}
Z.~Hu, S.~Shen, T.~Bonato, S.~Jeaugey, C.~Alexander, E.~Spada, J.~Dinan, J.~Hammond, and T.~Hoefler, ``{Demystifying NCCL: An In-Depth Analysis of GPU Communication Protocols and Algorithms},'' in \emph{2025 IEEE Symposium on High-Performance Interconnects}, 2025, pp. 48--59.

\bibitem{rccl2026}
{AMD}, ``{RCCL: ROCm Communication Collectives Library},'' \href{https://github.com/ROCm/rocm-systems/tree/develop/projects/rccl}{GitHub repository}, 2026, accessed October 6, 2026.

\bibitem{oneccl2026}
{UXL Foundation}, ``{oneCCL: oneAPI Collective Communications Library},'' \href{https://github.com/uxlfoundation/oneCCL}{GitHub repository}, 2026, accessed October 6, 2026.

\bibitem{ncclx2025}
M.~Si, P.~Balaji, Y.~Chen, C.-H. Chu, A.~Gangidi, S.~Hasan \emph{et~al.}, ``{Collective Communication for 100k+ GPUs},'' \emph{arXiv preprint arXiv:2510.20171}, 2025.

\bibitem{intelshmem2024}
A.~Brooks, P.~Marshall, D.~Ozog, M.~W. ur~Rahman, L.~Stewart, and R.~Tom, ``{Intel{\textregistered} SHMEM: GPU-initiated OpenSHMEM using SYCL},'' in \emph{SC24-W: Workshops of the International Conference for High Performance Computing, Networking, Storage and Analysis}, 2024, pp. 1288--1301.

\bibitem{mscclpp2026}
C.~Hwang, P.~Cheng, R.~Dathathri, A.~Jangda, S.~Maleki, M.~Musuvathi \emph{et~al.}, ``{MSCCL++: Rethinking GPU Communication Abstractions for AI Inference},'' in \emph{Proceedings of the 31st ACM International Conference on Architectural Support for Programming Languages and Operating Systems, Volume 2}, 2026, pp. 1201--1215.

\bibitem{ncclgin2025}
K.~Hamidouche, J.~Bachan, P.~Markthub, P.-J. Gootzen, E.~Agostini, S.~Jeaugey, A.~Shafi, G.~Theodorakis, and M.~G. Venkata, ``{GPU-Initiated Networking for NCCL},'' \emph{arXiv preprint arXiv:2511.15076}, 2025.

\bibitem{rcclgin2026}
{AMD}, ``{Use the RCCL Device API and GIN},'' \href{https://rocm.docs.amd.com/projects/rccl/en/develop/how-to/device-api-gin.html}{RCCL Documentation}, 2026, accessed October 6, 2026.

\bibitem{deepspeedmoe2022}
S.~Rajbhandari, C.~Li, Z.~Yao, M.~Zhang, R.~Y. Aminabadi, A.~A. Awan, J.~Rasley, and Y.~He, ``{DeepSpeed-MoE: Advancing Mixture-of-Experts Inference and Training to Power Next-Generation AI Scale},'' in \emph{Proceedings of the 39th International Conference on Machine Learning}, ser. Proceedings of Machine Learning Research, vol. 162, 2022, pp. 18\,332--18\,346.

\bibitem{tutel2023}
C.~Hwang, W.~Cui, Y.~Xiong, Z.~Yang, Z.~Liu, H.~Hu \emph{et~al.}, ``{Tutel: Adaptive Mixture-of-Experts at Scale},'' in \emph{Proceedings of Machine Learning and Systems}, vol.~5, 2023, pp. 269--287.

\bibitem{hybridep2026}
F.~Yu, T.~Liu, and K.~Sun, ``{Optimizing Communication for Mixture-of-Experts Training with Hybrid Expert Parallel},'' \href{https://developer.nvidia.com/blog/optimizing-communication-for-mixture-of-experts-training-with-hybrid-expert-parallel/}{NVIDIA Technical Blog}, 2026, accessed October 6, 2026.

\bibitem{mooncakeep2025}
{Mooncake Team}, ``{Mooncake-EP},'' \href{https://github.com/kvcache-ai/Mooncake}{GitHub repository}, 2025, accessed October 6, 2026.

\bibitem{nixl2025}
{NVIDIA}, ``{NIXL EP},'' \href{https://github.com/ai-dynamo/nixl/tree/main/examples/device/ep}{GitHub repository}, 2025, accessed October 6, 2026.

\bibitem{pplxkernels2025}
{Perplexity AI}, ``{pplx-kernels: Perplexity GPU Kernels},'' \href{https://github.com/perplexityai/pplx-kernels}{GitHub repository}, 2025, accessed October 6, 2026.

\bibitem{swiftep2026}
X.~Li, Y.~Liu, X.~Huang, Y.~Zhang, S.~Wang, S.~Wang \emph{et~al.}, ``{SwiftEP: Accelerating MoE Inference with Buffer Fusion and TMA Offloading},'' in \emph{23rd USENIX Symposium on Networked Systems Design and Implementation}.\hskip 1em plus 0.5em minus 0.4em\relax USENIX Association, 2026, pp. 1073--1089.

\bibitem{ncclep2026}
A.~Goldman, N.~Boker, M.~Sheraizin, N.~Admoni, A.~Polyakov, S.~Bhattacharya \emph{et~al.}, ``{NCCL EP: Towards a Unified Expert Parallel Communication API for NCCL},'' \emph{arXiv preprint arXiv:2603.13606}, 2026.

\bibitem{rocmdeepep2025}
{AMD}, ``{DeepEP for ROCm},'' \href{https://github.com/ROCm/DeepEP}{GitHub repository}, 2025, accessed October 6, 2026.

\bibitem{mori2025}
{ROCm}, ``{MORI: Modular RDMA Interface},'' \href{https://github.com/ROCm/mori}{GitHub repository}, 2025, accessed August 23, 2026.

\bibitem{moriep2025}
P.~Sun, A.~Luo, G.~Lei, L.~Jin, C.~Huang, D.~Wang \emph{et~al.}, ``{Practical, Fault-Robust Distributed Inference for DeepSeek on AMD MI300X},'' \href{https://rocm.blogs.amd.com/software-tools-optimization/wide-ep-deepseek/README.html}{ROCm Blogs}, Nov. 2025, accessed October 6, 2026.

\bibitem{oden2014ibverbs}
L.~Oden, H.~Fr{\"o}ning, and F.-J. Pfreundt, ``{Infiniband-Verbs on GPU: A Case Study of Controlling an Infiniband Network Device from the GPU},'' in \emph{2014 IEEE International Parallel \& Distributed Processing Symposium Workshops}, 2014, pp. 976--983.

\bibitem{gpurdma2016}
F.~Daoud, A.~Watad, and M.~Silberstein, ``{GPUrdma: GPU-side Library for High Performance Networking from GPU Kernels},'' in \emph{Proceedings of the 6th International Workshop on Runtime and Operating Systems for Supercomputers}, 2016, pp. 1--8.

\bibitem{gpudirectasync2018}
E.~Agostini, D.~Rossetti, and S.~Potluri, ``{GPUDirect Async: Exploring GPU Synchronous Communication Techniques for InfiniBand Clusters},'' \emph{Journal of Parallel and Distributed Computing}, vol. 114, pp. 28--45, 2018.

\bibitem{gputn2017}
M.~LeBeane, K.~Hamidouche, B.~Benton, M.~Breternitz, S.~K. Reinhardt, and L.~K. John, ``{GPU Triggered Networking for Intra-Kernel Communications},'' in \emph{Proceedings of the International Conference for High Performance Computing, Networking, Storage and Analysis}, 2017, pp. 1--12.

\bibitem{gpucentric2017}
S.~Potluri, A.~Goswami, D.~Rossetti, C.~J. Newburn, M.~G. Venkata, and N.~Imam, ``{GPU-Centric Communication on NVIDIA GPU Clusters with InfiniBand: A Case Study with OpenSHMEM},'' in \emph{2017 IEEE 24th International Conference on High Performance Computing}, 2017, pp. 253--262.

\bibitem{rocshmem2026}
{AMD}, ``{rocSHMEM: ROCm OpenSHMEM},'' \href{https://github.com/ROCm/rocm-systems/tree/develop/projects/rocshmem}{GitHub repository}, 2026, accessed October 4, 2026.

\bibitem{uep2026}
Z.~Mao, Y.~Zhang, C.~Cui, Z.~Huang, K.~You, Z.~Chen \emph{et~al.}, ``{UEP: Portable Expert-Parallel Communication},'' in \emph{20th USENIX Symposium on Operating Systems Design and Implementation (OSDI 26)}.\hskip 1em plus 0.5em minus 0.4em\relax USENIX Association, 2026, pp. 1107--1123.

\bibitem{fabriclib2026}
N.~Licker, K.~Hu, V.~Zaytsev, and L.~Chen, ``{fabric-lib: RDMA Point-to-Point Communication for LLM Systems},'' in \emph{Proceedings of Machine Learning and Systems}, vol.~8, 2026, pp. 169--185.

\bibitem{gpu_callbacks2010}
J.~A. Stuart, M.~Cox, and J.~D. Owens, ``{GPU-to-CPU Callbacks},'' in \emph{Euro-Par 2010 Parallel Processing Workshops}, ser. Lecture Notes in Computer Science.\hskip 1em plus 0.5em minus 0.4em\relax Springer, 2011, vol. 6586, pp. 365--372.

\bibitem{gpufs2013}
M.~Silberstein, B.~Ford, I.~Keidar, and E.~Witchel, ``{GPUfs: Integrating a File System with GPUs},'' in \emph{Proceedings of the 18th International Conference on Architectural Support for Programming Languages and Operating Systems}, 2013, pp. 485--498.

\bibitem{genesys2018}
J.~Vesel{\'y}, A.~Basu, A.~Bhattacharjee, G.~H. Loh, M.~Oskin, and S.~K. Reinhardt, ``{Generic System Calls for GPUs},'' in \emph{Proceedings of the 45th Annual International Symposium on Computer Architecture}, 2018, pp. 843--856.

\bibitem{openmpreverseoffload2023}
J.~Huber and J.~Chesterfield, ``{OpenMP Reverse Offloading Using Shared Memory Remote Procedure Calls},'' in \emph{OpenMP: Advanced Task-Based, Device and Compiler Programming}, ser. Lecture Notes in Computer Science.\hskip 1em plus 0.5em minus 0.4em\relax Springer Nature Switzerland, 2023, vol. 14114, pp. 226--238.

\bibitem{llvmlibcgpu2023}
J.~Huber, ``{The LLVM C Library for GPUs},'' \href{https://llvm.org/devmtg/2023-10/slides/techtalks/Huber-LibCforGPUs.pdf}{2023 LLVM Developers' Meeting}, 2023.

\bibitem{flydsl2026}
{ROCm}, ``{FlyDSL: Flexible Layout Python DSL},'' \href{https://github.com/ROCm/FlyDSL}{GitHub repository}, 2026, accessed October 14, 2026.

\bibitem{triton2019}
P.~Tillet, H.~T. Kung, and D.~Cox, ``{Triton: An Intermediate Language and Compiler for Tiled Neural Network Computations},'' in \emph{Proceedings of the 3rd ACM SIGPLAN International Workshop on Machine Learning and Programming Languages (MAPL)}, 2019, pp. 10--19.

\bibitem{rdmacore2026}
{Linux RDMA}, ``{rdma-core: RDMA Core Userspace Libraries and Daemons},'' \href{https://github.com/linux-rdma/rdma-core}{GitHub repository}, 2026, includes libibverbs.

\bibitem{libfabric2015}
P.~Grun, S.~Hefty, S.~Sur, D.~Goodell, R.~D. Russell, H.~Pritchard, and J.~M. Squyres, ``{A Brief Introduction to the OpenFabrics Interfaces -- A New Network API for Maximizing High Performance Application Efficiency},'' in \emph{Proceedings of the IEEE 23rd Annual Symposium on High-Performance Interconnects}, 2015, pp. 34--39.

\bibitem{disseminationbarrier1988}
D.~Hensgen, R.~Finkel, and U.~Manber, ``{Two Algorithms for Barrier Synchronization},'' \emph{International Journal of Parallel Programming}, vol.~17, no.~1, pp. 1--17, 1988.

\bibitem{vllm2023}
W.~Kwon, Z.~Li, S.~Zhuang, Y.~Sheng, L.~Zheng, C.~H. Yu, J.~E. Gonzalez, H.~Zhang, and I.~Stoica, ``{Efficient Memory Management for Large Language Model Serving with PagedAttention},'' in \emph{Proceedings of the ACM Symposium on Operating Systems Principles}, 2023, pp. 611--626.

\bibitem{glm5report2026}
{GLM-5-Team}, A.~Zeng, X.~Lv, Z.~Hou, Z.~Du, Q.~Zheng \emph{et~al.}, ``{GLM-5: from Vibe Coding to Agentic Engineering},'' \emph{arXiv preprint arXiv:2602.15763}, 2026.

\bibitem{glm53modelcard2026}
{Z.ai}, ``{GLM-5.3},'' \href{https://huggingface.co/zai-org/GLM-5.3}{Hugging Face model card}, Aug. 2026, accessed October 29, 2026.

\bibitem{cobbe2021gsm8k}
K.~Cobbe, V.~Kosaraju, M.~Bavarian, M.~Chen, H.~Jun, L.~Kaiser \emph{et~al.}, ``{Training Verifiers to Solve Math Word Problems},'' \emph{arXiv preprint arXiv:2110.14168}, 2021.

\bibitem{kamradt2023niah}
G.~Kamradt, ``{Needle In A Haystack: Pressure Testing LLMs},'' \href{https://github.com/gkamradt/LLMTest_NeedleInAHaystack}{GitHub repository}, 2023, accessed October 27, 2026.

\bibitem{nvfp42025}
{NVIDIA}, F.~Abecassis, A.~Agrusa, D.~Ahn, J.~Alben, S.~Alborghetti \emph{et~al.}, ``{Pretraining Large Language Models with NVFP4},'' \emph{arXiv preprint arXiv:2509.25149}, 2025.

\bibitem{latentmoe2026}
V.~Elango, N.~Bhatia, R.~Waleffe, R.~Shafipour, T.~Asida, A.~Khattar \emph{et~al.}, ``{LatentMoE: Toward Optimal Accuracy per FLOP and Parameter in Mixture of Experts},'' \emph{arXiv preprint arXiv:2601.18089}, 2026.

\bibitem{mlaas2022}
Q.~Weng, W.~Xiao, Y.~Yu, W.~Wang, C.~Wang, J.~He, Y.~Li, L.~Zhang, W.~Lin, and Y.~Ding, ``{MLaaS in the Wild: Workload Analysis and Scheduling in Large-Scale Heterogeneous GPU Clusters},'' in \emph{Proceedings of the 19th USENIX Symposium on Networked Systems Design and Implementation}, 2022, pp. 945--960.

\bibitem{acme2024}
Q.~Hu, Z.~Ye, Z.~Wang, G.~Wang, M.~Zhang, Q.~Chen \emph{et~al.}, ``{Characterization of Large Language Model Development in the Datacenter},'' in \emph{Proceedings of the 21st USENIX Symposium on Networked Systems Design and Implementation}, 2024, pp. 709--729.

\end{thebibliography}
